\documentclass[letterpaper,11pt]{article}

\pdfoutput=1

\usepackage{hyperref}

\hypersetup{
    colorlinks=true,       
    linkcolor=blue,          
    citecolor=blue,        
    filecolor=blue,      
    urlcolor=blue           
}

\usepackage{float}
                     
\usepackage{braket,slashed,bm}
\usepackage{array,multirow}
\usepackage[normalem]{ulem}
\usepackage{xcolor,cancel,youngtab}

\usepackage[T1]{fontenc} 
\usepackage{enumitem}
\usepackage{mathrsfs}
\usepackage{booktabs}
\usepackage{adjustbox}
\usepackage{mathtools}

\usepackage{soul}

\usepackage{tikz}
\usetikzlibrary{arrows,decorations.pathmorphing,backgrounds,positioning,fit,petri,automata,shadows,calendar,mindmap,
decorations.markings,calc}

\definecolor{labelkey}{rgb}{0,0.5,0.0}

\usepackage{xspace}
\usepackage{slashed}
\usepackage{array,multirow}
\usepackage{graphicx}
\usepackage{graphics}
\usepackage{epsfig}
\usepackage{amsfonts}
\usepackage{amsmath}
\usepackage{amssymb}
\usepackage{dsfont}
\usepackage{color}
\usepackage{xcolor}
\usepackage{cite}
\usepackage{pdflscape}
\usepackage{pifont}
\usepackage{pgfplots}
\usepackage{tikz} 
\usepackage{pgfplotstable}
\usepackage{bbold}

\newcommand{\hc}{\mathrm{h.c.}}
\newcommand{\ep}{\epsilon}

\newcommand{\al}{\alpha}
\newcommand{\bt}{\beta}
\newcommand{\g}{\gamma}

\newcommand{\simu}{\sigma^{\mu\nu}}

\newcommand{\vL}{\ensuremath{\mathcal{L}}}

\usepackage{bm}  %

\newcommand{\beq}{\begin{equation}}
\newcommand{\eeq}{\end{equation}}
\newcommand{\be}{\begin{equation}}
\newcommand{\ee}{\end{equation}}
\newcommand{\bea}{\begin{eqnarray}}
\newcommand{\eea}{\end{eqnarray}}
\newcommand{\ben}{\begin{eqnarray*}}
\newcommand{\een}{\end{eqnarray*}}

\renewcommand{\vec}[1]{{\mathbf #1}} 

\newcommand{\bma}{\begin{pmatrix}}
\newcommand{\ema}{\end{pmatrix}}

\def\lixo#1{}

\def\slashchar#1{\setbox0=\hbox{$#1$}           
  \dimen0=\wd0                                    
  \setbox1=\hbox{/} \dimen1=\wd1                  
  \ifdim\dimen0>\dimen1                           
    \rlap{\hbox to \dimen0{\hfil/\hfil}}            
    #1                                             
  \else                                          
    \rlap{\hbox to \dimen1{\hfil$#1$\hfil}}        
    /                                           
 \fi}                                           %

\newcommand{\dslash}[1]{#1 \llap{/\kern-0.5pt}}
\newcommand{\Dslash}[1]{#1 \llap{/\kern+1.5pt}}
\newcommand{\DDslash}[1]{#1 \llap{/\kern+2.3pt}}
\newcommand{\dslashh}[1]{#1 \llap{/\kern+1pt}}

\newcommand{\nn}{\nonumber}

\newcommand{\textoverline}[1]{$\overline{\mbox{#1}}$}

\definecolor{cadmiumgreen}{rgb}{0.0, 0.42, 0.24}
\definecolor{darkpastelgreen}{rgb}{0.01, 0.75, 0.24}
\definecolor{darkspringgreen}{rgb}{0.09, 0.45, 0.27}
\definecolor{forestgreen(web)}{rgb}{0.13, 0.55, 0.13}
\definecolor{forestgreen(traditional)}{rgb}{0.0, 0.27, 0.13}
\definecolor{cobalt}{rgb}{0.0, 0.28, 0.67}
\definecolor{darkblue}{rgb}{0.0, 0.0, 0.75}
\definecolor{darkred}{rgb}{0.55, 0.0, 0.0}
\definecolor{palatinatepurple}{rgb}{0.41, 0.16, 0.38}
\definecolor{burntorange}{rgb}{0.8, 0.33, 0.0}

\begin{document}

\begin{titlepage}

\begin{flushright}
\phantom{ LA-UR-xx-xxxx}
\end{flushright}

\vspace{2.0cm}

\begin{center}
{\LARGE  \bf 
Light sterile neutrinos and lepton-number-violating kaon decays in effective field theory\\ 
}
\vspace{2cm}

{\large \bf  Guanghui Zhou$^{a,b}$\footnote{g.zhou@uva.nl}} 
\vspace{0.5cm}

\vspace{0.25cm}

\vspace{0.25cm}
{\large 
	$^a$ 
	{\it 
	Institute for Theoretical Physics Amsterdam and Delta Institute for Theoretical Physics, University of Amsterdam, Science Park 904, 1098 XH Amsterdam, The Netherlands}}

\vspace{0.25cm}
{\large 
	$^b$ 
	{\it 
	Nikhef, Theory Group, Science Park 105, 1098 XG, Amsterdam, The Netherlands}}

\end{center}

\vspace{0.2cm}

\begin{abstract}
\vspace{0.1cm}

We investigate lepton-number-violating decays  $K^\mp \rightarrow \pi^\pm l^\mp l^\mp$ in the presence of sterile neutrinos. We consider minimal interactions  with Standard-Model fields through Yukawa couplings as well as higher-dimensional operators in the framework of the neutrino-extended Standard Model Effective Field Theory. We use $SU(3)$ chiral perturbation theory to match to mesonic interactions and compute the lepton-number-violating decay rate in terms of the neutrino masses and the Wilson coefficients of higher-dimensional operators. For neutrinos that can be produced on-shell, the decay rates are highly enhanced and higher-dimensional interactions can be probed up to very high scales around $ \mathcal{O}$(30) TeV.

\end{abstract}

\vfill
\end{titlepage}

\tableofcontents

\section{Introduction}
Neutrino oscillations have demonstrated that neutrinos are massive particles, which cannot be explained within the Standard Model (SM) of particle physics in its original form. Some Beyond-the-SM (BSM) physics is required explain the origin of neutrino masses. Arguably the simplest modification is to add a right-handed gauge-singlet neutrino field (a sterile neutrino) to the SM. This field can couple to the left-handed neutrino (active neutrino) field  and the Higgs field through Yukawa interactions, generating a neutrino Dirac mass term in exact analogue to other fermions. It is possible to add a Majorana mass term for the sterile neutrino as neither Lorentz nor gauge symmetry forbids this. Such a term violates Lepton number (L), an accidental symmetry of the SM, by two units. The combination of a Majorana mass term and the Yukawa interactions results in neutrinos that are  Majorana mass eigenstates, and the presence of lepton-number-violating (LNV) processes such as neutrinoless double beta decay \cite{Cirigliano:2017djv,Cirigliano:2018yza,Dekens:2020ttz,Pas:1999fc,Pas:2000vn,Blennow:2010th,Mitra:2011qr,Li:2011ss,Barea:2015zfa,Giunti:2015kza} or LNV kaon decays \cite{Chun:2019nwi,Zhang:2021wjj,Abad:1984gh,Ng:1978ij,Godbole:2020doo,Cvetic:2010rw,Ivanov:2004ch,Dib:2000wm,Littenberg:2000fg,Littenberg:1991ek}. 

In recent years, an effective field theory (EFT) approach has been developed for neutrinoless double beta decay based on the framework of the SMEFT \cite{Cirigliano:2017djv,Cirigliano:2018yza} or its neutrino-extended version $\nu$SMEFT \cite{Dekens:2020ttz}. In $\nu$SMEFT, the SM is extended with higher-dimensional SU$(2)_L$$\times$U$(1)_Y$ gauge-invariant operators consisting of SM fields and gauge-singlet neutrinos $\nu_R$. Higher-dimensional operators are more and more suppressed by powers of $v/\Lambda$ where $\Lambda$ is the scale of BSM physics and they are listed in Refs. \cite{Li:2020gnx,Li:2020xlh,Li:2020tsi,Li:2021tsq,Liao:2016qyd}. 
LNV operators begin at the renormalizable level, the $\nu_R$ Majorana mass term, while all the remaining LNV operators have odd dimension $\geq 5$ \cite{Kobach:2016ami}. In this work, we extend this approach to the mesonic analogue of neutrinoless double decay, the LNV kaon decays $K^\mp \rightarrow \pi^\pm l^\mp l^\mp$ ($l=e, \mu$). We note that this process has been studied in the SMEFT in Refs.~ \cite{Liao:2019gex,Liao:2020roy}.

The description of low-energy LNV processes depends on the mass scale of sterile neutrinos. If neutrinos are heavier than the electroweak (EW) scale $v\simeq 246$ GeV, they can be integrated out and their low-energy signature is captured by local gauge-invariant effective SMEFT operators with odd dimension. When neutrino masses are below the EW scale, the operators of $\nu$SMEFT are evolved to the EW scale  and heavy SM particles (top, W, Z, Higgs) are integrated out to match to a Fermi-like EFT extended with $\nu_R$ fields that obeys $SU(3)_{c} \times U(1)_{em}$ gauge symmetry.  If the sterile neutrinos are heavier than $\Lambda_\chi\simeq 1 $ GeV,  they can be integrated out before matching to a chiral EFT Lagrangian. We obtain LNV dimension-9 operators, which involve two up-type quarks, two down-type quarks and two charged leptons.  These operators induce short-distance LNV contributions and were systematic studied in Refs.~\cite{Liao:2019gex,Dekens:2020ttz}. Sterile neutrinos with masses below  $\Lambda_\chi$ are active degrees of freedom at hadronic scales. We apply $SU(3)$ chiral EFT extended with sterile neutrinos to describe LNV kaon decays in this mass regime. Such neutrinos can be looked for in many different experiments, ranging from oscillation to beta-decay to collider experiments \cite{Bolton:2019pcu,Bischer:2019ttk,Bolton:2020ncv,Li:2020wxi, Dasgupta:2021ies, Dekens:2021qch,Cirigliano:2021peb,Liao:2021qfj}. 

 The current experimental upper bounds on the LNV branching ratios of charged kaons are very stringent ($5.3\times 10^{-11}$ and $4.2\times 10^{-11}$ \cite{NA62:2019eax,NA62:2022tte} for $K^- \rightarrow \pi^+ e^- e^-$ and $K^- \rightarrow \pi^+ \mu^- \mu^-$, respectively). Nevertheless, unlike the case for neutrinoless double beta decay these bounds are too weak to set meaningful constraints on the BSM scale $\Lambda$ for the exchange of virtual sterile neutrinos . However, if a sterile neutrino can be produced on shell, the LNV decay rate is significantly enhanced due to the small width of the sterile neutrino \cite{Dib:2000wm}. We apply the narrow-width approximation to modify the decay amplitude and  use the resonance to constrain the neutrino mixing angles and the BSM scale. Similar ideas were used for other kinds of  LNV  decays including $D, D_s, B, B_s, B_c$, and $\tau $ decays, see Refs.~\cite{Cvetic:2010rw,Atre:2009rg,Helo:2010cw,Gribanov:2001vv,Mejia-Guisao:2017gqp,Cvetic:2017vwl,Milanes:2016rzr} for discussions on the minimal scenario without higher-dimensional operators and Ref. \cite{Godbole:2020jqw} on the left-right symmetric model.

The paper is organized as follows.  We discuss the $\nu$SMEFT framework in Sec. \ref{sec:smeft} and then we give the expressions for the LNV kaon decay amplitudes in terms of the Wilson coefficients (WCs), neutrino masses and hadronic low-energy constants for the long- and short-distance contributions  in Sec. \ref{chiral}. The phase space integral and resonance are introduced in Sec. \ref{sec:int} and then we discuss the phenomenology of two scenarios and give the limits on the WCs of several \textoverline{dim-6} operators in Sec. \ref{sec:pheno}.  We conclude this work in Sec. \ref{sec:conclusion}. App. \ref{app:matching}  gives the matching conditions to connect  operators before and after the EW symmetry breaking (EWSB). The dim-9 interactionss proportional to $m_l$ or $m_q$ are discessed in App. \ref{app:matchd9} and we give the decay expressions of the sterile neutrino in App. \ref{app:decay}. 
\section{The operators in $\nu$SMEFT up to dimension seven  }\label{sec:smeft}
In this work, we denote the dimensions of $\nu$SMEFT operators by \textoverline{dim-n}  with $n=5,6,7$ and the dimensions of the operators after EWSB by dim-n with $n=3, 6, 7, 9$. At the BSM physics scale $\Lambda\gg v$, the relevant Lagrangian can be written as
\begin{eqnarray}\label{eq:smeft}
\mathcal L &=&  \mathcal L_{SM} - \left[ \frac{1}{2} \bar \nu^c_{R} \,\bar M_R \nu_{R} +\bar L \tilde H Y_\nu \nu_R + \rm{h.c.}\right]\nn \\
&&+  \mathcal L^{(\bar 5)}_{\nu_L}+  \mathcal L^{(\bar 5)}_{\nu_R}+  \mathcal L^{(\bar 6)}_{\nu_L}+  \mathcal L^{(\bar 6)}_{\nu_R} +   \mathcal L^{(\bar 7)}_{\nu_L} +   \mathcal L^{(\bar 7)}_{\nu_R}\,,
\end{eqnarray}
where $\mathcal L_{SM} $ is the Lagrangian from the SM, $L$ denotes the lepton doublet and $H$ is the Higgs doublet with $\tilde{H}=i\tau_2 H^*$. In unitary gauge, we can write
 \begin{equation}
 H = \frac{v}{\sqrt{2}}  \left(\begin{array}{c}
 0 \\
 1 + \frac{h}{v}
 \end{array} \right)\,,
 \end{equation}
where $v=246$ GeV is the Higgs vacuum expectation value (vev), $h$ is the Higgs field. $\nu_R$ is a  column vector of $n$ sterile neutrinos, $Y_\nu$ is a $3\times n$ matrix of Yukawa couplings and $\bar{M}_R$  is a complex symmetric Majorana mass matrix of type $n\times n$ that violates lepton number by two units. We  choose a basis where the charged leptons $e^i_{L,R}$ and quarks $u^i_{L,R}$ and $d^i_{R}$ are in mass eigenstates with $i$=1,2,3. While for the left handed down-type quarks, we have $d^i_L=V^{ij}d^{j,\rm mass}_L$ with $V$ being the CKM matrix. The charge conjugate field of $\Psi$ is $\Psi^c=C\bar{\Psi}^T$, where the charge conjugation matrix $C$ is $-i\gamma^2\gamma^0$ and it satisfies $C=-C^{-1}=-C^T=-C^\dagger$.   For chiral fields we have $\Psi_{L,R}^c = (\Psi_{L,R})^c =  C \overline{\Psi_{L,R}}^T= P_{R,L} \Psi^c$, with $P_{R,L}=(1\pm\gamma_5)/2$.

 \begin{figure}
	\includegraphics[scale=0.23]{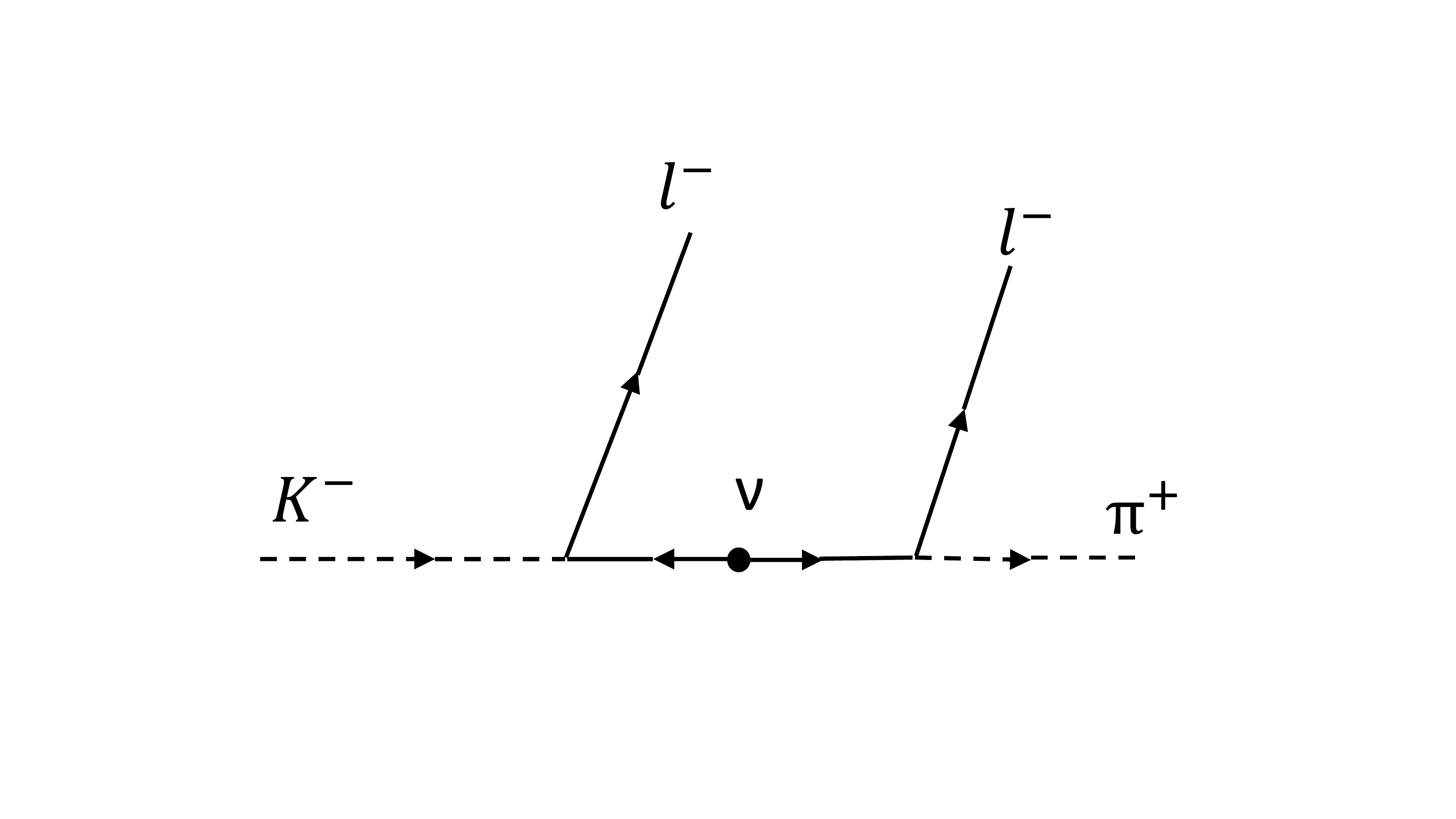}
	\includegraphics[scale=0.23]{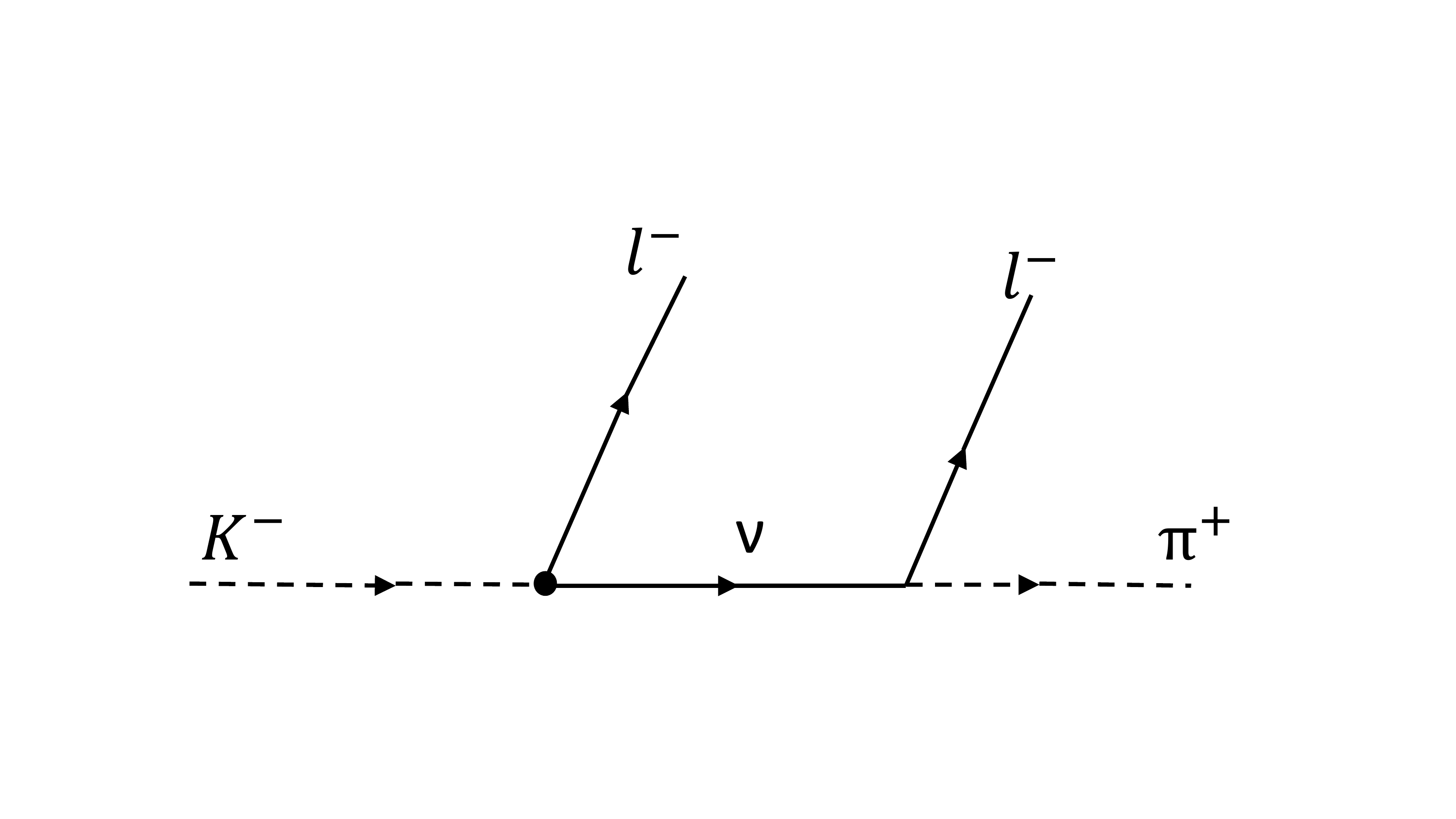}	
	\includegraphics[scale=0.23]{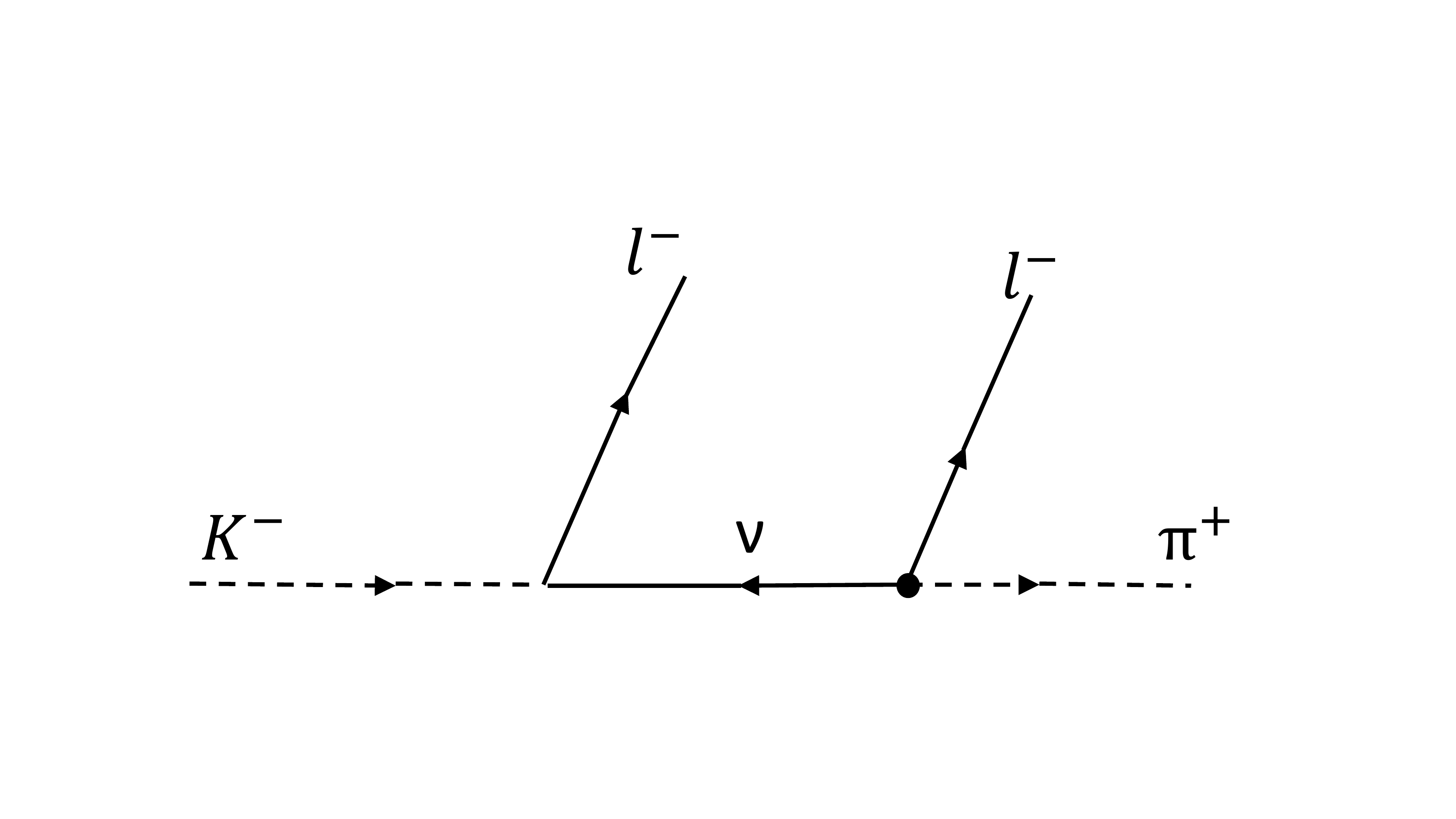}
	\includegraphics[scale=0.23]{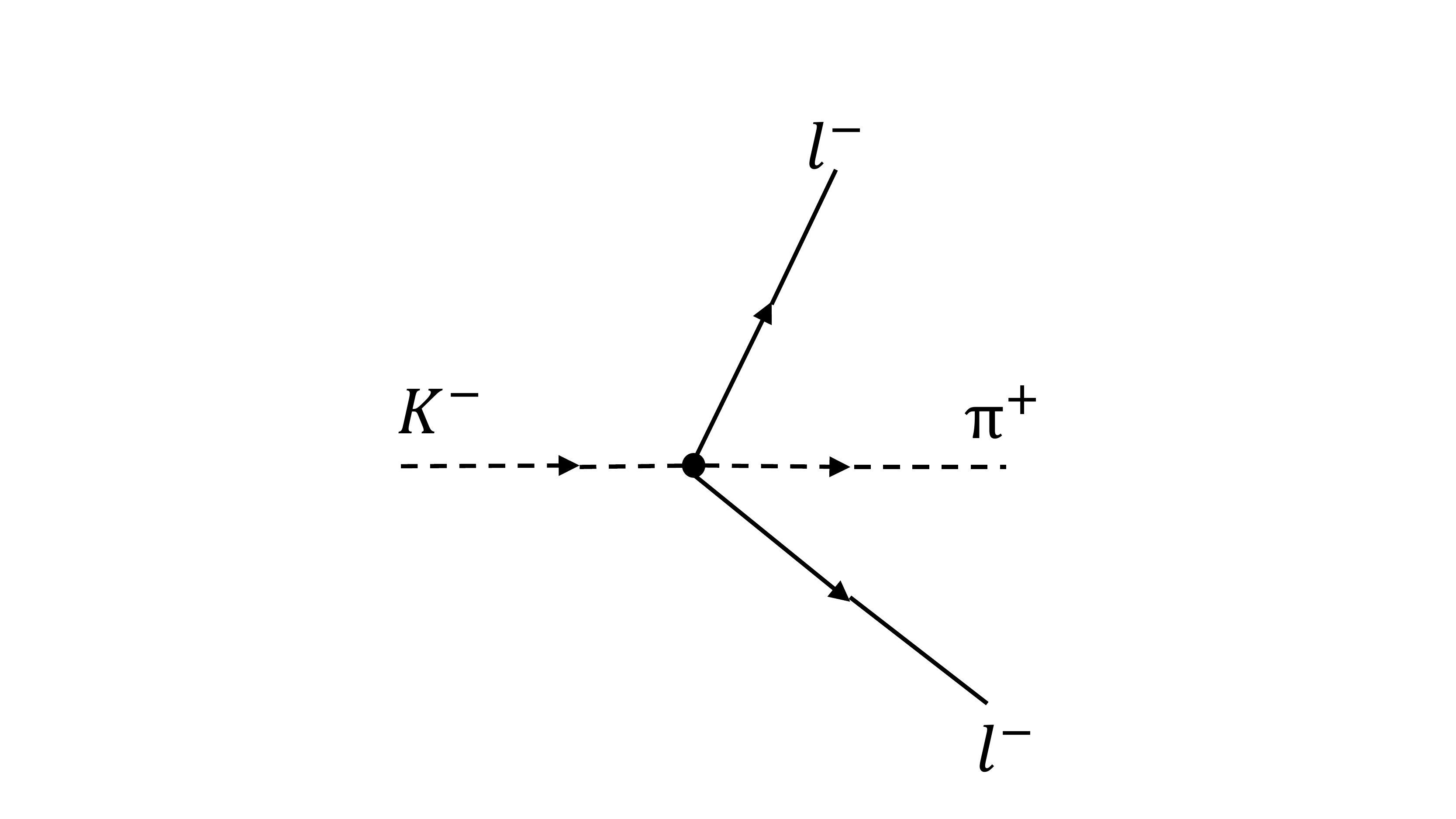}
	\caption{Possible Feynman diagrams for $K^-\rightarrow \pi^+l^-l^-$ and the black blob denotes LNV interactions.}\label{diagram}
\end{figure} 

 We present the possible Feynman diagrams relevant with $K^-\rightarrow \pi^+l^-l^-$ in Fig. \ref{diagram}. The first three accomplish LNV through the exchange of a light neutrino (long-distance contribution). They can be divided into two parts, one for the leptonic decay of $K^-$ and the other for the leptonic decay of $\pi^-$. To get the hadronic operators that induce leptonic decay of mesons,  we need quark-level operators involving an up quark, a down (strange) quark, a charged lepton and a neutrino, which are at least dim-6 after EWSB. The fourth diagram contains no neutrino and we call it the short-distance contribution. It may be induced by integrating heavy neutrinos out or arise  from \textoverline{dim-7} operators without neutrinos (see Eqs. (\ref{eq:d9match}) and \eqref{match9}).

There are two relevant operators at \textoverline{dim-5}
\be
\mathcal L^{(\bar 5)}_{\nu_L} = \ep_{kl}\ep_{mn}(L_k^T\, C^{( 5)}\,CL_m )H_l H_n\,,\qquad  \mathcal L^{(\bar 5)}_{\nu_R}=- \bar \nu^c_{R} \,\bar M_R^{(5)} \nu_{R} H^\dagger H\,,
\ee
which after EWSB  contribute to the Majorana mass terms for active and sterile neutrinos, respectively. There is also a \textoverline{dim-5} transition dipole operator which is not relevant with this work  and is ignored.  We mainly focus on operators that involve only one neutrino  appearing at  \textoverline{dim-6} and \textoverline{dim-7}. The operators of $\mathcal L^{(\bar 6)}_{\nu_L}$ and  $\mathcal L^{(\bar 6)}_{\nu_R}$ are \textoverline{dim-6} and involve a left-handed neutrino and a sterile neutrino, respectively. We list them in tables \ref{tab:O6L} and \ref{tab:O6R}.  Similarly we  list the  \textoverline{dim-7} operators in tables \ref{tab:O7L} and \ref{tab:O7R}.
{\renewcommand{\arraystretch}{1.3}\begin{table}[t]\small
		\center
		\begin{tabular}{||c|c||c|c||c|c||}
			\hline Class $1$& $\psi^2 H X$ & Class $2$ & $\psi^2 H^2 D$        & Class $3$ &  $\psi^4 $\\
			\hline
			$\mathcal O_{eW}^{(6)}$  &        $(\bar L  \sigma^{\mu\nu} e) \tau^I H W_{\mu\nu}^I$ & 
			$ \mathcal O_{H L\,3}^{(6)}$ &   $(H^\dag i\overleftrightarrow{D}^I_\mu H)(\bar L \tau^I \gamma^\mu L)$   &
			$\mathcal O_{LeQu\,1}^{(6)}$&  $(\bar L ^j e) \epsilon_{jk} (\bar Q^k u)$  \\
			$\mathcal O_{uW}^{(6)}$  &        $(\bar Q \sigma^{\mu\nu} u) \tau^I \widetilde H \, W_{\mu\nu}^I$ &
			$\mathcal O_{H Q\,3}^{(6)}$      & $(H^\dag i\overleftrightarrow{D}^I_\mu H)(\bar Q \tau^I \gamma^\mu Q)$ &
			$\mathcal O_{Lequ\,3}^{(6)}$ &    $(\bar L^j \sigma_{\mu\nu} e) \epsilon_{jk} (\bar Q^k \sigma^{\mu\nu} u)$\\
			$\mathcal O_{dW}^{(6)}$ & $(\bar Q \sigma^{\mu\nu} d) \tau^I H\, W_{\mu\nu}^I$ &
			$\mathcal O_{H u d}^{(6)}$   & $i(\widetilde H ^\dag D_\mu H)(\bar u \gamma^\mu d)$ &
			$\mathcal O_{LQ\,3}^{(6)}$     & $(\bar L \gamma^\mu \tau^I L)(\bar Q \gamma_\mu \tau^I Q)$ \\
			&  & & & $\mathcal O_{LedQ}^{(6)}$ & $(\bar L^j e)(\bar d Q^{j})$ \\\hline
			\hline 
		\end{tabular}
		\caption{LNC \textoverline{dim-6} operators \cite{Grzadkowski:2010es} of  $\mathcal L^{(\bar 6)}_{\nu_L}$  that affect LNV kaon decays at tree level.}   \label{tab:O6L}
\end{table}}

{\renewcommand{\arraystretch}{1.3}
	\begin{table}[t]\small
		\center
		\begin{tabular}{||c|c||c|c||}
			\hline Class $1$& $\psi^2 H^4$  & Class $5$ &  $\psi^4 D$\\
			\hline
			$\mathcal O^{(7)}_{LH}$  & $\ep_{ij}\ep_{mn}(L_i^TCL_m )H_j H_n (H^\dagger H)$ & $\mathcal O^{(7)}_{LL\bar d u D\,1}$&  $\ep_{ij} (\bar d \g_\mu u)(L_i^T C (D^\mu L)_j)$  \\\hline
			Class $2$&  $\psi^2 H^2 D^2$ & Class $6$ & $\psi^4 H$\\\hline
			$\mathcal O^{(7)}_{LHD\,1}$  & $\ep_{ij}\ep_{mn}(L_i^TC (D_\mu L)_j )H_m  (D^\mu H)_n$ & $\mathcal O^{(7)}_{L e u \bar d  H}$   & $\ep_{ij}( L_i^T C\g_\mu e)(\bar d\g^\mu  u )H_j$ \\
			$\mathcal O^{(7)}_{LHD\,2}$  & $\ep_{im}\ep_{jn}(L_i^TC (D_\mu L)_j )H_m  (D^\mu H)_n$  & $\mathcal O^{(7)}_{LL Q\bar d  H\,1}$  &  $\ep_{ij}\ep_{mn}(\bar d L_i)(Q_j^TC L_m )H_n$ \\
			\cline{1-2} Class $3$& $\psi^2 H^3 D$  & $\mathcal O^{(7)}_{LL Q\bar d  H\,2}$  &  $\ep_{im}\ep_{jn}(\bar d L_i)(Q_j^TC L_m )H_n$\\\cline{1-2} 
			$\mathcal O^{(7)}_{LHDe}$ & $\ep_{ij}\ep_{mn}(L_i^TC \g_\mu e )H_j H_m  (D^\mu H)_n$ & $\mathcal O^{(7)}_{LL \bar Q u  H}$    &$\ep_{ij}(\bar Q_m u)(L_m^TC L_i )H_j$\\
			\cline{1-2} \cline{1-2} Class $4$& $\psi^2 H^2 X $ & &\\\cline{1-2} 
			$\mathcal O^{(7)}_{LHW}$  & $\ep_{ij} (\ep\tau^I)_{mn} g(L_i^TC \simu L_m ) H_j H_n W^I_{\mu\nu}$ & & \\
			\hline
		\end{tabular}
		\caption{LNV \textoverline{dim-7}operators \cite{Lehman:2014jma} of  $\mathcal L^{(\bar 7)}_{\nu_L}$  that affect LNV kaon decays at tree level.  
		} \label{tab:O7L}
\end{table}}

{\renewcommand{\arraystretch}{1.3}\begin{table}[t]\small
		\center
		\begin{tabular}{||c|c||c|c||}
			\hline Class $1$& $\psi^2 H^3$  & Class $4$ &  $\psi^4 $\\
			\hline
			$\mathcal{O}^{(6)}_{L\nu H}$ & $(\bar{L}\nu_R)\tilde{H}(H^\dagger H)$ & $\mathcal{O}^{(6)}_{du\nu e}$ & $ (\bar{d}\gamma^\mu u)(\bar{\nu_R }\gamma_\mu e)$  \\ \cline{1-2}
			Class $2$&  $\psi^2 H^2 D$ &  $\mathcal{O}^{(6)}_{Qu\nu L}$ & $(\bar{Q}u)(\bar{\nu}_RL)$  \\ \cline{1-2}
			$\mathcal{O}^{(6)}_{H\nu e}$ & $(\bar{\nu }_R\gamma^\mu e)({\tilde{H}}^\dagger i D_\mu H)$ & $\mathcal{O}^{(6)}_{L\nu Qd}$ & $(\bar{L}\nu_R )\epsilon(\bar{Q}d))$ \\ \cline{1-2}
			Class $3$ & $\psi^2 H^3 D$  & $\mathcal{O}^{(6)}_{LdQ\nu }$ & $(\bar{L}d)\epsilon(\bar{Q}\nu_R )$ \\ \cline{1-2}
			$\mathcal{O}^{(6)}_{\nu W}$ &$(\bar{L}\sigma_{\mu\nu}\nu_R )\tau^I\tilde{H}W^{I\mu\nu}$  & &\\
			\hline
		\end{tabular}
		\caption{LNC \textoverline{dim-6} operators \cite{Liao:2016qyd} of  $\mathcal L^{(\bar 6)}_{\nu_R}$  that affect LNV kaon decays at tree level.
		} \label{tab:O6R}
\end{table}}

{\renewcommand{\arraystretch}{1.3}\begin{table}[t]\small
		\center
		\begin{tabular}{||c|c||c|c||}
			\hline Class $1$& $\psi^2 H^4$  & Class $5$ &  $\psi^4 D$\\
			\hline
			$\mathcal{O}_{\nu H}^{(7)}$ &$ (\nu^T_R C\nu_R)(H^\dagger H)^2$ & $\mathcal{O}_{du\nu eD}^{(7)} $&$(\bar{d}\gamma_\mu u)(\nu^T_R CiD_\mu e)$  \\\cline{1-2}
			Class $2$&  $\psi^2 H^2 D^2$ &  $\mathcal{O}_{QL\nu uD}^{(7)}$&$(\bar{Q}\gamma_\mu L)(\nu^T_RCiD_\mu u)$\\\cline{1-2}
			$\mathcal{O}_{\nu eD}^{(7)}$&$\epsilon_{ij}(\nu^T_R CD_\mu e)(H^iD^\mu H^j)$ & $\mathcal{O}_{d\nu QLD}^{(7)}$&$\epsilon_{ij}(\bar{d}\gamma_\mu \nu_R)(Q^iCiD_\mu L^j)$ \\\hline
			Class $3$& $\psi^2 H^3 D$  & Class $6$    &$\psi^4H$ \\\hline
			$\mathcal{O}_{\nu L1}^{(7)}$& $\epsilon_{ij}(\nu^T_R C\gamma_\mu L^i) (iD^\mu H^j)(H^\dagger H)$ & $\mathcal{O}_{Q\nu QLH2}^{(7)}$&$\epsilon_{ij}(\bar{Q}\nu_R)(Q^iCL^j)H$\\\cline{1-2}
			Class $4$& $\psi^2 H^2 X $ & $\mathcal{O}_{dL\nu uH}^{(7)}$& $\epsilon_{ij}(\bar{d}L^i)(\nu^T_RCu)\tilde{H}^j$\\\cline{1-2} 
			$\mathcal{O}_{\nu eW}^{(7)}$&$(\epsilon\tau^I)_{ij}(\nu^T_RC\sigma^{\mu\nu}e)(H^iH^j)W^I_{\mu\nu}$ & $\mathcal{O}_{dQ\nu eH}^{(7)}$&$\epsilon_{ij}(\bar{d}Q^i)(\nu^T_R Ce)H^j $ \\
			& & $\mathcal{O}_{Qu\nu eH}^{(7)}$&$(\bar{Q}u)(\nu^T_R Ce)H$ \\
			& & $\mathcal{O}_{Qe\nu uH}^{(7)}$&$(\bar{Q}e)(\nu^T_R Cu)H$ \\
			\hline
		\end{tabular}
		\caption{LNV \textoverline{dim-7}operators \cite{Liao:2016qyd} of  $\mathcal L^{(\bar 7)}_{\nu_R}$  that affect LNV kaon decays at tree level.  
		} \label{tab:O7R}
\end{table}}

\subsection{The Lagrangian after EWSB}
After EWSB, heavy SM particles with masses above the EW scale are integrated out and we are left with a $SU(3)_c\times U(1)_{em}$-invariant EFT which we call LEFT. The Lagrangian of LEFT can be written as
\begin{eqnarray}\label{DL2lag}
\mathcal L_{} &=&  \mathcal L_{SM}-  \left[\frac{1}{2} \bar \nu^c_{L} \, M_L \nu_{L}   +  \frac{1}{2} \bar \nu^c_{R} \, M_R \nu_{R} +\bar \nu_L M_D\nu_R +\hc \right]\nn \\
&&+  \mathcal L^{(6)}_{\Delta L = 2}+  \mathcal L^{(6)}_{\Delta L = 0} +   \mathcal L^{(7)}_{\Delta L = 2} +   \mathcal L^{(7)}_{\Delta L =0 } + \mathcal L^{(9)}_{\Delta L = 2}\,,
\end{eqnarray}
where $\mathcal{L}_{SM}$ denotes operators of dim-4 and lower of  light SM fields and  we discuss the matching conditions  in App. \ref{app:matching}. Part of the operators in the second line are \cite{Dekens:2020ttz}
\bea
\mathcal L^{(6)}_{\Delta L = 2}& =& \frac{2 G_F}{\sqrt{2}} \Bigg\{ 
\bar u_L \gamma^\mu d_L \left[  \bar e_{R}  \gamma_\mu C^{(6)}_{\textrm{VL}} \,  \nu^c_{L}+ \bar e_{L}  \gamma_\mu \bar C^{(6)}_{\textrm{VL}} \,  \nu^c_{R} \right]+
\bar u_R \gamma^\mu d_R \left[\bar e_{R}\,  \gamma_\mu  C^{(6)}_{\textrm{VR}} \,\nu_{L}^c+\bar e_{L}\,  \gamma_\mu  \bar C^{(6)}_{\textrm{VR}} \,\nu^c_{R}  \right]\nn \\
& & +
\bar u_L  d_R \left[ \bar e_{L}\, C^{(6)}_{ \textrm{SR}}  \nu^c_{L} +\bar e_{R}\, \bar C^{(6)}_{ \textrm{SR}}  \nu^c_{R} \right]+ 
\bar u_R  d_L \left[ \bar e_{L} \, C^{(6)}_{ \textrm{SL}}    \nu_{L}^c + \bar e_{R} \, \bar C^{(6)}_{ \textrm{SL}}    \nu_{R}^c \right]\nn \\
&&+  \bar u_L \sigma^{\mu\nu} d_R\,  \bar e_{L}  \sigma_{\mu\nu} C^{(6)}_{ \textrm{T}} \, \nu_{L}^c+  \bar u_R \sigma^{\mu\nu} d_L\,  \bar e_{R}  \sigma_{\mu\nu} \bar C^{(6)}_{ \textrm{T}} \, \nu_{R}^c
\Bigg\}  +{\rm h.c.}\label{eq:lowenergy6_l2}\eea  
\bea
\mathcal L^{(6)}_{\Delta L = 0}& =& \frac{2 G_F}{\sqrt{2}} \Bigg\{ 
\bar u_L \gamma^\mu d_L \left[  \bar e_{L}  \gamma_\mu c^{(6)}_{\textrm{VL}} \,  \nu_{L}+ \bar e_{R}  \gamma_\mu \bar c^{(6)}_{\textrm{VL}} \,  \nu_{R} \right]+
\bar u_R \gamma^\mu d_R \left[\bar e_{L}\,  \gamma_\mu  c^{(6)}_{\textrm{VR}} \,\nu_{L}+\bar e_{R}\,  \gamma_\mu  \bar c^{(6)}_{\textrm{VR}} \,\nu_{R}  \right]\nn \\
& & +
\bar u_L  d_R \left[ \bar e_{R}\, c^{(6)}_{ \textrm{SR}}  \nu_{L} +\bar e_{L}\, \bar c^{(6)}_{ \textrm{SR}}  \nu_{R} \right]+ 
\bar u_R  d_L \left[ \bar e_{R} \, c^{(6)}_{ \textrm{SL}}    \nu_{L} + \bar e_{L} \, \bar c^{(6)}_{ \textrm{SL}}    \nu_{R} \right]\nn \\
&&+  \bar u_R \sigma^{\mu\nu} d_L\,  \bar e_{R}  \sigma_{\mu\nu} c^{(6)}_{ \textrm{T}} \, \nu_{L}+  \bar u_L \sigma^{\mu\nu} d_R\,  \bar e_{L}  \sigma_{\mu\nu} \bar c^{(6)}_{ \textrm{T}} \, \nu_{R}
\Bigg\}  +{\rm h.c.} \label{eq:lowenergy6_l0}
\eea
\bea
\mathcal L^{(7)}_{\Delta L = 2} &=& \frac{2 G_F}{\sqrt{2} v} \Bigg\{ 
\bar u_L \gamma^\mu d_L \left[ \bar e_{L} \, C^{(7)}_{\textrm{VL}} \,  i \overleftrightarrow{D}_\mu  \nu_{L}^c+ \bar e_{R} \, \bar C^{(7)}_{\textrm{VL}} \,  i \overleftrightarrow{D}_\mu  \nu_{R}^c \right] \nn \\
&&+
\bar u_R \gamma^\mu d_R \left[ \bar e_{L} \, C^{(7)}_{\textrm{VR}}\, i \overleftrightarrow{D}_\mu  \nu^c_{L}+ \bar e_{R} \, \bar C^{(7)}_{\textrm{VR}}\, i \overleftrightarrow{D}_\mu  \nu^c_{R} \right] \nn \\
&&+  \bar u_L \sigma^{\mu\nu} d_R\, \bar e_L \bar C_{\rm TR}^{(7)} \overleftarrow \partial_\mu \gamma_\nu \nu_R^c+  \bar u_R \sigma^{\mu\nu} d_L\, \bar e_L  \bar C_{\rm TL}^{(7)}\gamma_\nu\partial_\mu \nu_R^c
\Bigg\}  +{\rm h.c.}\label{eq:lowenergy7}\eea
\bea 
\mathcal L^{(7)}_{\Delta L = 0} &=& \frac{2 G_F}{\sqrt{2} v} \Bigg\{ 
\bar u_L \gamma^\mu d_L \left[ \bar e_{R} \, c^{(7)}_{\textrm{VL}} \,  i \overleftrightarrow{D}_\mu  \nu_{L} +\bar e_{L} \, \bar c^{(7)}_{\textrm{VL}} \,  i \overleftrightarrow{D}_\mu  \nu_{R} \right] \nn \\
&&+
\bar u_R \gamma^\mu d_R \left[ \bar e_{R} \, c^{(7)}_{\textrm{VR}}\, i \overleftrightarrow{D}_\mu  \nu_{L}   +\bar e_{L} \, \bar c^{(7)}_{\textrm{VR}}\, i \overleftrightarrow{D}_\mu  \nu_{R}\right]\nn \\
&&+
\bar u_L \sigma^{\mu\nu} d_R\, \partial_\mu\left(\bar e_L c_{\rm TR}^{(7)}   \gamma_\nu \nu_L\right)+  \bar u_R \sigma^{\mu\nu} d_L\, \partial_\mu\left(\bar e_L c_{\rm TL}^{(7)}   \gamma_\nu \nu_L\right)
\Bigg\}  +{\rm h.c.}\label{eq:lowenergy7b}
\eea
where $\overleftrightarrow D_\mu = D_\mu - \overleftarrow D_\mu$ and each Wilson coefficient carries indices $ijkl$ with $i$ = $u$  denoting the up quark, $j=d,s$ for the down quark and strange quark,  $k=\{e,\mu\}$ for electron and muon,    $l=\{1,2,3\}$ for active neutrinos and $l=\{1,\dots,n\}$ for sterile neutrinos. 

Dim-9 operators are induced at the electroweak scale and, when sterile neutrinos have a mass above the chiral-symmetry-breaking scale $\Lambda_\chi$, at the sterile neutrino mass threshold. We thus list all possible $SU(3)_c\times U(1)_{em}$ invariant operators  
\bea \label{eq:Lag9}
\vL^{(9)}_{\Delta L =2} = \frac{1}{v^5}\sum_i\bigg[\left( C^{(9)}_{i\, \rm R}\, \bar e_R C \bar e^T_{R} + C^{(9)}_{i\, \rm L}\, \bar e_L C \bar e^T_{L} \right)  \, O_i +  C^{(9)}_i\bar e\g_\mu\g_5  C \bar e^T\, O_i^\mu\bigg]\,,
\eea
where $O_i$ and $O_i^\mu$ contain four quarks  and they are Lorentz scalars and vectors, respectively. In general there should also be Lorentz  tensors, but in this work we only consider the case when both of the two outgoing leptons are electrons or muons and thus the tensor operators vanish. 
The scalar operators are \cite{Prezeau:2003xn,Graesser:2016bpz,Liao:2019gex}
\bea\label{eq:d9scalar}
O_ 1  &=&  \bar{u}_L^\alpha  \gamma_\mu  d_L^\alpha \ \bar{u}_L^\beta  \gamma^\mu  s_L^\beta\,, \qquad O^\prime_ 1  =  \bar{u}_R^\alpha  \gamma_\mu  d_R^\alpha \ \bar{u}_R^\beta  \gamma^\mu  s_R^\beta     \nn
\,\,, 
\\
O_ 2  &=&  \bar{u}_R^\alpha   d_L^\alpha \  \bar{u}_R^\beta   s_L^\beta\,, \qquad \qquad     O^\prime_ 2  =  \bar{u}_L^\alpha   d_R^\alpha \  \bar{u}_L^\beta   s_R^\beta \nn
\,\,,\\
O_ 3  &=&  \bar{u}_R^\alpha   d_L^\beta \  \bar{u}_R^\beta   s_L^\alpha\,, \qquad \qquad    O^\prime_ 3  =  \bar{u}_L^\alpha   d_R^\beta \  \bar{u}_L^\beta   s_R^\alpha \nn
\,\,,\\
O_ 4  &=&  \bar{u}_R^\alpha    d_L^\alpha \  \bar{u}_L^\beta   s_R^\beta\,, \qquad \qquad O^\prime_ 4  =  \bar{u}_L^\alpha    d_R^\alpha \  \bar{u}_R^\beta    s_L^\beta   \nn
\,\,,\\
O_ 5  &=&  \bar{u}_R^\alpha    d_L^\beta \  \bar{u}_L^\beta    s_R^\alpha\,, \qquad\qquad O^\prime_ 5  =  \bar{u}_L^\alpha    d_R^\beta \  \bar{u}_R^\beta    s_L^\alpha \,,\label{dim9scalar}
\eea
where $\al$, $\bt$ are color indices and a summation over them is implied.  We  get $O_i'$ by changing the parity of $O_i$ . For the  vector operators, we have  \cite{Graesser:2016bpz,Liao:2019gex}
\bea
O^\mu_ {6, \rm udus}  &=&  \bar{u}_L^\alpha  \gamma^\mu  d_L^\alpha \ \bar{u}_L^\beta    s_R^\beta\,, \qquad O^{\mu\prime}_ {6, \rm udus}  =  \bar{u}_R^\alpha  \gamma^\mu  d_R^\alpha \ \bar{u}_R^\beta    s_L^\beta    \nn
\,\,, 
\\
O^\mu_ {6, \rm usud}  &=&  \bar{u}_L^\alpha  \gamma^\mu  s_L^\alpha \ \bar{u}_L^\beta    d_R^\beta\,, \qquad O^{\mu\prime}_ {6, \rm usud}  = \bar{u}_R^\alpha  \gamma^\mu  s_R^\alpha \ \bar{u}_R^\beta    d_L^\beta     \nn
\,\,, 
\\
O^\mu_ {7,\rm udus}  &=&  \bar{u}_L^\alpha  \gamma^\mu  d_L^\beta \ \bar{u}_L^\beta    s_R^\alpha\,, \qquad O^{\mu\prime}_ {7, \rm udus}  =  \bar{u}_R^\alpha  \gamma^\mu  d_R^\beta \ \bar{u}_R^\beta    s_L^\alpha    \nn
\,\,, 
\\
O^\mu_ {7, \rm usud}  &=&  \bar{u}_L^\alpha  \gamma^\mu  s_L^\beta \ \bar{u}_L^\beta    d_R^\alpha\,, \qquad O^{\mu\prime}_ {7,\rm usud}  =\bar{u}_R^\alpha  \gamma^\mu  s_R^\beta \ \bar{u}_R^\beta    d_L^\alpha     \nn
\,\,, 
\\
O^\mu_ {8, \rm udus}  &=&  \bar{u}_L^\alpha  \gamma^\mu  d_L^\alpha \ \bar{u}_R^\beta    s_L^\beta\,, \qquad O^{\mu\prime}_ {8, \rm udus}  =  \bar{u}_R^\alpha  \gamma^\mu  d_R^\alpha \ \bar{u}_L^\beta    s_R^\beta \nn
\,\,, 
\\
O^\mu_ {8, \rm usud}  &=&  \bar{u}_L^\alpha  \gamma^\mu  s_L^\alpha \ \bar{u}_R^\beta    d_L^\beta\,, \qquad O^{\mu\prime}_ {8,\rm usud}  = \bar{u}_R^\alpha  \gamma^\mu  s_R^\alpha \ \bar{u}_L^\beta    d_R^\beta \nn
\,\,, 
\\
O^\mu_ {9, \rm udus}  &=&  \bar{u}_L^\alpha  \gamma^\mu  d_L^\beta \ \bar{u}_R^\beta    s_L^\alpha\,, \qquad O^{\mu\prime}_ {9,\rm udus}  =  \bar{u}_R^\alpha  \gamma^\mu  d_R^\beta \ \bar{u}_L^\beta    s_R^\alpha \nn\,\,,
\\
O^\mu_ {9,\rm usud}  &=&  \bar{u}_L^\alpha  \gamma^\mu  s_L^\beta \ \bar{u}_R^\beta    d_L^\alpha\,, \qquad O^{\mu\prime}_ {9, \rm usud}  = \bar{u}_R^\alpha  \gamma^\mu  s_R^\beta \ \bar{u}_L^\beta    d_R^\alpha \,.
\label{eq:d9vector}
\eea
The higher-dimensional operators are evolved from some BSM scale $\mu =\Lambda$ to the EW scale $\mu=m_W$ and then to the QCD scale $\Lambda_\chi$. The renormalization group equations  due to one-loop QCD effects are discussed in Ref. \cite{Dekens:2020ttz, Liao:2020roy}.  Their effect on the limits of new physics energy scale $\Lambda$ discussed in Sec. \ref{sec:pheno} is less than 30\% and is neglected in this work.

\subsection{Rotation to the neutrino mass basis}
We  write the neutrino mass terms as 
 \bea
\mathcal L_m = -\frac{1}{2} \bar N^c M_\nu N +{\rm h.c.}\,,\qquad M_\nu = \bma M_L &M_D^*\\M_D^\dagger&M_R^\dagger \ema \,,
\eea
where $N = (\nu_L,\, \nu_R^c)^T$ and $M_\nu$ is a $\bar{n}\times \bar{n}$ symmetric matrix, with $\bar{n}=3+n$. The mass matrix can be diagonalized by a  $\bar{n}\times \bar{n}$ unitary matrix, $U$, 
\bea\label{Mdiag}
U^T M_\nu U =m_\nu = {\rm diag}(m_1,\dots , m_{3+n})\,, \qquad N = U N_m\,,
\eea
where $m_1,\dots ,m_{\bar{n}}$ are real and non-negative. We define  $\nu = N_m +N_m^c = \nu^c$, which are the Majorana mass eigenstates. The kinetic and mass terms of the neutrinos become
\bea
\mathcal L_\nu = \frac{1}{2} \bar \nu i\slashed \partial \nu -\frac{1}{2} \bar \nu^{ } m_\nu \nu\,.
\eea
 The relation between neutrinos in mass basis and flavor basis is 
\bea
\nu_L = P_L(P U) \nu \,,\qquad \nu_L^c =P_R (P U^*) \nu\,,\nn\\
\nu_R =P_R (P_s U^*) \nu \,,\qquad \nu_R^c = P_L(P_s U) \nu\,,
\eea
where $P$ and $P_s$ are $3\times \bar{n}$ and $n \times \bar{n}$ projector matrices, respectively
\be
P = \begin{pmatrix}\mathcal I_{3\times 3} & 0_{3 \times n}  \end{pmatrix}\,,\qquad
P_s = \begin{pmatrix} 0_{n\times 3} & \mathcal I_{n \times n}  \end{pmatrix}\, .
\ee
In the mass basis, we can write the higher-dimensional operators in a remarkably compact form
\bea\label{6final}
\mathcal L^{(6)}& =& \frac{2 G_F}{\sqrt{2}} \Bigg\{ 
\bar u_L \gamma^\mu d_L \left[  \bar e_{R}  \gamma_\mu C^{(6)}_{\textrm{VLR}} \,  \nu+ \bar e_{L}  \gamma_\mu  C^{(6)}_{\textrm{VLL}} \,  \nu \right]+
\bar u_R \gamma^\mu d_R \left[\bar e_{R}\,  \gamma_\mu  C^{(6)}_{\textrm{VRR}} \,\nu+\bar e_{L}\,  \gamma_\mu   C^{(6)}_{\textrm{VRL}} \,\nu  \right]\nn\\
& & +
\bar u_L  d_R \left[ \bar e_{L}\, C^{(6)}_{ \textrm{SRR}}  \nu +\bar e_{R}\,  C^{(6)}_{ \textrm{SRL}}  \nu \right]+ 
\bar u_R  d_L \left[ \bar e_{L} \, C^{(6)}_{ \textrm{SLR}}    \nu + \bar e_{R} \,  C^{(6)}_{ \textrm{SLL}}    \nu \right]\nn\\
&&+  \bar u_L \sigma^{\mu\nu} d_R\,  \bar e_{L}  \sigma_{\mu\nu} C^{(6)}_{ \textrm{TRR}} \, \nu+  \bar u_R \sigma^{\mu\nu} d_L\,  \bar e_{R}  \sigma_{\mu\nu}  C^{(6)}_{ \textrm{TLL}} \, \nu
\Bigg\}  +{\rm h.c.}\eea
and for the dim-7 operators we have
\bea\label{7final}
\mathcal L^{(7)}&=& \frac{2 G_F}{\sqrt{2} v} \Bigg\{ 
\bar u_L \gamma^\mu d_L \left[ \bar e_{L} \, C^{(7)}_{\textrm{VLR}} \,  i \overleftrightarrow{D}_\mu  \nu+\bar e_{R} \,  C^{(7)}_{\textrm{VLL}} \,  i \overleftrightarrow{D}_\mu  \nu \right] \nn\\
&&+
\bar u_R \gamma^\mu d_R \left[ \bar e_{L} \, C^{(7)}_{\textrm{VRR}}\, i \overleftrightarrow{D}_\mu  \nu+ \bar e_{R} \,  C^{(7)}_{\textrm{VRL}}\, i \overleftrightarrow{D}_\mu  \nu \right] \nn\\
&&+  \bar u_L \sigma^{\mu\nu} d_R\, \bar e_L  C_{\rm TR1}^{(7)} \overleftarrow D_\mu \gamma_\nu \nu
+  \bar u_R \sigma^{\mu\nu} d_L\, \bar e_L  C_{\rm TL1}^{(7)}\gamma_\nu \partial_\mu \nu \nn\\
&&+  \bar u_L \sigma^{\mu\nu} d_R\, D_\mu\left(\bar e_L C_{\rm TR2}^{(7)}   \gamma_\nu \nu\right)
+  \bar u_R \sigma^{\mu\nu} d_L\,D_\mu\left(\bar e_L C_{\rm TL2}^{(7)}   \gamma_\nu \nu\right)
\Bigg\}  +{\rm h.c.}
\eea
The Wilson coefficients satisfy the following relation
 \begin{align}\label{redefC6}
 C_{\rm VLR}^{(6)} &= 	  C_{\rm VL}^{(6)}PU^*	+\bar c_{\rm VL}^{(6)}P_s U^*\,,\qquad 	&C_{\rm VRR}^{(6)} &= C_{\rm VR}^{(6)}PU^*+\bar c_{\rm VR}^{(6)}P_s U^*\,,\nn\\
 C_{\rm VLL}^{(6)} &= \bar C_{\rm VL}^{(6)}P_sU	+     c_{\rm VL}^{(6)}P U\,,	\qquad 	&C_{\rm VRL}^{(6)} &= \bar C_{\rm VR}^{(6)}P_sU+ c_{\rm VR}^{(6)}P U\,,\nn\\
 C_{\rm SLR}^{(6)} &= 	  C_{\rm SL}^{(6)}PU^*	+\bar c_{\rm SL}^{(6)}P_s U^*\,,\qquad  &C_{\rm SRR}^{(6)} &= C_{\rm SR}^{(6)}PU^*+\bar c_{\rm SR}^{(6)}P_s U^*\,,\nn\\
 C_{\rm SLL}^{(6)} &=  \bar C_{\rm SL}^{(6)}P_sU+ c_{\rm SL}^{(6)}P U\,,		\qquad  &C_{\rm SRL}^{(6)} &= \bar C_{\rm SR}^{(6)}P_sU+ c_{\rm SR}^{(6)}P U\,,\nn\\
 C_{\rm TLL}^{(6)} &=  \bar C_{\rm T}^{(6)}P_sU+ c_{\rm T}^{(6)}P U\,,		\qquad 	&C_{\rm TRR}^{(6)} &=  C_{\rm T}^{(6)}PU^*+ \bar c_{\rm T}^{(6)}P_s U^*\,,\nn\\
 C_{\rm VLL}^{(7)} &= c_{\rm VL}^{(7)}PU+\bar C_{\rm VL}^{(7)}P_s U\,,\qquad &C_{\rm VRL}^{(7)} &= c_{\rm VR}^{(7)}PU+\bar C_{\rm VR}^{(7)}P_s U\,,\nn\\
 C_{\rm VLR}^{(7)} &= C_{\rm VL}^{(7)}PU^*+\bar c_{\rm VL}^{(7)}P_s U^*\,,\qquad &C_{\rm VRR}^{(7)} &=  C_{\rm VR}^{(7)}PU^*+\bar c_{\rm VR}^{(7)}P_s U^*\,,\nn\\
 C_{\rm TL1}^{(7)} &= \bar C_{\rm TL}^{(7)}P_s U\,,\qquad &C_{\rm TL2}^{(7)} &= c_{\rm TL}^{(7)}PU\,,\nn\\
 C_{\rm TR1}^{(7)} &= \bar C_{\rm TR}^{(7)}P_sU\,,\qquad &C_{\rm TR2}^{(7)} &= c_{\rm TR}^{(7)}PU\,.
 \end{align}
 These coefficients carry flavor indices $ijkl$ where $i= u$ denotes the up quark, $j=d, s$ indicate the down quark and strange quark, $k= e,\mu$ labels the generation of charged lepton and   $l = \{1, ...,\bar{n}\} $ denote  neutrinos in the mass basis.  
The rotation has no influence on the dim-9 operators as they contain no neutrino fields. We focus mainly on the operators in Eq.~\eqref{6final} as the operators in Eq.~\eqref{7final} are relatively suppressed by $m_\pi/v$ or $v/\Lambda$.

\subsection{Integrating out sterile   neutrinos when  $\Lambda_\chi  < m_{\nu} \leq v$}\label{dim9}
To integrate out heavy neutrinos above the chiral-symmetry-breaking scale $\Lambda_\chi$, we write the Lagrangian containing heavy neutrinos as
\bea
\vL_H=\sum_{i=1}^{n_H} \bigg[\frac{1}{2} \bar \nu_i i\slashed\partial \nu_i -\frac{1}{2}\bar \nu_i m_{\nu_i} \nu_i +\mathcal J_i \nu_i\bigg]\,,
\eea
where neutrinos are in the mass eigenstates and we sum over $n_H$ heavy neutrinos whose masses satisfy $\Lambda_\chi  < m_{\nu_i} \leq v$. $\mathcal{J}_i$ contains the interactions for the $i$-th neutrino. 
By using the equation of motion we integrate out the heavy neutrinos and  get the following effective Lagrangian

\bea\label{eq:d9match}
\vL_{eff}&\simeq&\frac{1}{2m^2_{\nu_i}}\mathcal  J_{i} (i \slashed \partial +m_{\nu_i})C\mathcal J^T_{i} \,,\nn\\
\mathcal J_i &=&J_{ud,i}+J_{us,i}+\bar J_{ud,i}+\bar J_{us,i}\,,\nn\\
J_{ud,i} &\simeq& \frac{1}{v^2} 
\bigg[\bar u_L \gamma^\mu d_L \left[  \bar e_{R}  \gamma_\mu C^{(6)}_{\textrm{VLR}} + \bar e_{L}  \gamma_\mu  C^{(6)}_{\textrm{VLL}} \right]+
\bar u_R \gamma^\mu d_R \left[\bar e_{R}\,  \gamma_\mu  C^{(6)}_{\textrm{VRR}} +\bar e_{L}\,  \gamma_\mu   C^{(6)}_{\textrm{VRL}}   \right]\nn\\
&  +&
\bar u_L  d_R \left[ \bar e_{L}\, C^{(6)}_{ \textrm{SRR}}  +\bar e_{R}\,  C^{(6)}_{ \textrm{SRL}}\right]+ 
\bar u_R  d_L \left[ \bar e_{L} \, C^{(6)}_{ \textrm{SLR}}  + \bar e_{R} \,  C^{(6)}_{ \textrm{SLL}}  \right]\nn\\
& +&  \bar u_L \sigma^{\mu\nu} d_R\,  \bar e_{L}  \sigma_{\mu\nu} C^{(6)}_{ \textrm{TRR}} +  \bar u_R \sigma^{\mu\nu} d_L\,  \bar e_{R}  \sigma_{\mu\nu}  C^{(6)}_{ \textrm{TLL}} \bigg]_i\,,\nn\\
J_{us,i}&=&J_{ud,i}|_{d\rightarrow s}\,,
\eea
where $\bar J_i$ is the hermitian conjugate of $J_i$, a sum over $i$ is implied and by transposing the leptonic part of $\mathcal J_i$ we get $\mathcal J^T_i$.  The interactions relevant with $K^-\rightarrow \pi^+ l^-l^-$ are
\bea
\vL^{(9)}_{H}&\simeq&\sum_{i=1}^{n_H}\frac{1}{m^2_{\nu_i}}  J_{us,i} (i \slashed \partial +m_{\nu_i})C J^T_{ud,i} \,,
\eea
 which contain two kinds of  terms, one proportional to $\frac{1}{m_{\nu_i}}$ and the other proportional to $\frac{1}{m^2_{\nu_i}}$ with an additional derivative. Here we give the matching conditions for terms of the first kind.  We find for the scalar dim-9 operators 
\bea\label{eq:matchingscalar9}
C^{(9)}_{1\,R} &=& -v C_{\rm VLR,us}^{(6)}\bar m_\nu^{-1} C_{\rm VLR,ud}^{(6)\, T}\,,\qquad C^{(9)\prime }_{1\,R} = -v C_{\rm VRR,us}^{(6)}\bar m_\nu^{-1} C_{\rm VRR,ud}^{(6)\, T}\,,\nn\\
C^{(9)}_{2\,R} &=& v C_{\rm SLL,us}^{(6)}\bar m_\nu^{-1} C_{\rm SLL,ud}^{(6)\, T}-16v C_{\rm TLL,us}^{(6)}\bar m_\nu^{-1} C_{\rm TLL,ud}^{(6)\, T}\,,\qquad  C^{(9)\prime }_{2\,R} = v C_{\rm SRL,us}^{(6)}\bar m_\nu^{-1} C_{\rm SRL,ud}^{(6)\, T}\,,\nn\\
C^{(9)}_{3\,R} &=&-32v C_{\rm TLL,us}^{(6)}\bar m_\nu^{-1} C_{\rm TLL,ud}^{(6)\, T}\,,\qquad C^{(9)\prime }_{3\,R}=0\,,\nn\\
C^{(9)}_{4\,R} &=&v C_{\rm SRL,us}^{(6)}\bar m_\nu^{-1} C_{\rm SLL,ud}^{(6)\, T}\,,\qquad  C^{(9)\prime}_{4\,R} =v C_{\rm SLL,us}^{(6)}\bar m_\nu^{-1} C_{\rm SRL,ud}^{(6)\, T}  \,,\nn\\
C^{(9)}_{5\,R} &=&2v C_{\rm VRR,us}^{(6)}\bar m_\nu^{-1} C_{\rm VLR,ud}^{(6)\, T}\,,\qquad C^{(9)\prime}_{5\,R} =2v C_{\rm VLR,us}^{(6)}\bar m_\nu^{-1} C_{\rm VRR,ud}^{(6)\, T} \,,
\eea
where  $C_{\rm XXX,us}^{(6)}$ and $C_{\rm XXX,ud}^{(6)}$  are the coefficients of dim-6 operators in Eq.~\eqref{6final} involving  a strange quark and a down quark, respectively.  We  get the matching conditions for the $C^{(9)}_{i\,L}$ operators  via the replacement
\bea\label{LtoR}
C^{(9)}_{i\,R} & \to &C^{(9)\prime }_{i\,L}\,, \qquad C^{(9) \prime}_{i\,R} \to C^{(9) }_{i\,L}\,,\qquad \, \nn\\
C_{\rm ALL}^{(6)}&\leftrightarrow & C^{(6)}_{\rm ARR}\,,\qquad C_{\rm ARL}^{(6)}\leftrightarrow C^{(6)}_{\rm ALR}\,,\qquad A\in \{S,V,T\}\,.
\eea
The matching contributions for the vector dim-9 operators are given by
\bea\label{dim93}
C_{6,\rm usud}^{(9)}&=&\frac{v}{2}\left( C_{\rm VLR,us}^{(6)} m_{\nu_i}^{-1} C_{\rm SRR,ud}^{(6)\, T}-C_{\rm VLL,us}^{(6)} m_{\nu_i}^{-1} C_{\rm SRL,ud}^{(6)\, T}\right)\nn\\
&&+\frac{1}{2}C_{7,\rm usud}^{(9)}\,,\nn\\
C_{6,\rm udus}^{(9)}&=&\frac{v}{2}\left( C_{\rm SRR,us}^{(6)} m_{\nu_i}^{-1} C_{\rm VLR,ud}^{(6)\, T}-C_{\rm SRL,us}^{(6)} m_{\nu_i}^{-1} C_{\rm VLL,ud}^{(6)\, T}\right)\nn\\
&&+\frac{1}{2}C_{7,\rm udus}^{(9)}\,,\nn\\
C_{7,\rm usud}^{(9)}&=&4 v  C_{\rm VLR,us}^{(6)} m_{\nu_i}^{-1} C_{\rm TRR,ud}^{(6)\, T}\,,\nn\\
C_{7,\rm udus}^{(9)}&=&4 v  C_{\rm TRR,us}^{(6)} m_{\nu_i}^{-1} C_{\rm VLR,ud}^{(6)\, T}\,,\nn\\
C_{6,\rm usud}^{(9)\prime}&=&\frac{v}{2}\left( C_{\rm VRR,us}^{(6)} m_{\nu_i}^{-1} C_{\rm SLR,ud}^{(6)\, T}-C_{\rm VRL,us}^{(6)} m_{\nu_i}^{-1} C_{\rm SLL,ud}^{(6)\, T}\right)\nn\\
&&+\frac{1}{2}C_{7,\rm usud}^{(9)\prime}\,,\nn\\
C_{6,\rm udus}^{(9)\prime}&=&\frac{v}{2}\left( C_{\rm SLR,us}^{(6)} m_{\nu_i}^{-1} C_{\rm VRR,ud}^{(6)\, T}-C_{\rm SLL,us}^{(6)} m_{\nu_i}^{-1} C_{\rm VRL,ud}^{(6)\, T}\right)\nn\\
&&+\frac{1}{2}C_{7,\rm udus}^{(9)\prime}\,,\nn\\
C_{7,\rm usud}^{(9)\prime}&=&-4 v  C_{\rm VRL,us}^{(6)} m_{\nu_i}^{-1} C_{\rm TLL,ud}^{(6)\, T}\,,\nn\\
C_{7,\rm udus}^{(9)\prime}&=&-4 v  C_{\rm TLL,us}^{(6)} m_{\nu_i}^{-1} C_{\rm VRL,ud}^{(6)\, T}\,,\nn\\
C_{8,\rm usud}^{(9)}&=&\frac{v}{2}\left( C_{\rm VLR,us}^{(6)} m_{\nu_i}^{-1} C_{\rm SLR,ud}^{(6)\, T}-C_{\rm VLL,us}^{(6)} m_{\nu_i}^{-1} C_{\rm SLL,ud}^{(6)\, T}\right)\nn\\
&&+\frac{1}{2}C_{9,\rm udus}^{(9)}\,,\nn\\
C_{8,\rm udus}^{(9)}&=&\frac{v}{2}\left( C_{\rm SLR,us}^{(6)} m_{\nu_i}^{-1} C_{\rm VLR,ud}^{(6)\, T}-C_{\rm SLL,us}^{(6)} m_{\nu_i}^{-1} C_{\rm VLL,ud}^{(6)\, T}\right)\nn\\
&&+\frac{1}{2}C_{9,\rm usud}^{(9)}\,,\nn\\
C_{9,\rm usud}^{(9)}&=&4 v  C_{\rm TLL,us}^{(6)} m_{\nu_i}^{-1} C_{\rm VLL,ud}^{(6)\, T}\,,\nn\\
C_{9,\rm udus}^{(9)}&=&4 v  C_{\rm VLL,us}^{(6)} m_{\nu_i}^{-1} C_{\rm TLL,ud}^{(6)\, T}\,,\nn\\
C_{8,\rm usud}^{(9)\prime}&=&\frac{v}{2}\left( C_{\rm VRR,us}^{(6)} m_{\nu_i}^{-1} C_{\rm SRR,ud}^{(6)\, T}-C_{\rm VRL,us}^{(6)} m_{\nu_i}^{-1} C_{\rm SRL,ud}^{(6)\, T}\right)\nn\\
&&+\frac{1}{2}C_{9,\rm udus}^{(9)\prime}\,,\nn\\
C_{8,\rm udus}^{(9)\prime}&=&\frac{v}{2}\left( C_{\rm SRR,us}^{(6)} m_{\nu_i}^{-1} C_{\rm VRR,ud}^{(6)\, T}-C_{\rm SRL,us}^{(6)} m_{\nu_i}^{-1} C_{\rm VRL,ud}^{(6)\, T}\right)\nn\\
&&+\frac{1}{2}C_{9,\rm usud}^{(9)\prime}\,,\nn\\
C_{9,\rm usud}^{(9)\prime}&=&-4 v  C_{\rm TRR,us}^{(6)} m_{\nu_i}^{-1} C_{\rm VRR,ud}^{(6)\, T}\,,\nn\\
C_{9,\rm udus}^{(9)\prime}&=&-4 v  C_{\rm VRR,us}^{(6)} m_{\nu_i}^{-1} C_{\rm TRR,ud}^{(6)\, T}\,.
\eea
The matching conditions for  terms proportional to $\frac{1}{m^2_{\nu_i}}$  are given in App. \ref{app:matchd9}. 
\section{ Chiral perturbation theory with  sterile neutrinos}\label{chiral}
\subsection{The case of light sterile neutrinos}
Below the GeV scale, quarks and gluons can not be used as degrees of freedom due to the strong dynamics. We thus use the framework of chiral perturbation theory ($\chi$PT) \cite{Weinberg:1978kz} to connect hadronic physics with the higher-dimensional operators. When the neutrino mass is below GeV scale it is an explicit degree of freedom in $\chi$PT. We use the external source method and write the QCD Lagrangian in Eq. \eqref{6final}  as 
\bea
\mathcal L_{qq} &=& \bar q i \Dslash \partial q+ \bar q\bigg\{ l^\mu\gamma_\mu P_L +r^\mu \gamma_\mu P_R \nn\\
&&- (M+s+ip)P_L - (M+s-ip)P_R + t^{\mu\nu}_L\sigma_{\mu\nu} P_L  + t^{\mu\nu}_R\sigma_{\mu\nu} P_R\big\}q\,,
\eea 
where $q= (u,d,s)^T$ is the triplet of quark fields, and $M=\mathrm{diag} (m_u,\,m_d,\,m_s)$ is a diagonal $3\times 3$ matrix for the quark masses. The external sources can be read from Eq.~\eqref{6final} 
\bea\label{sources}
s+ip &=& -\frac{2 G_F}{\sqrt{2}}\left\{\lambda_i\left(\bar e_L C_{\rm SLR}^{(6)} \nu + \bar e_R C_{\rm SLL}^{(6)} \nu\right)_{}+\left(\lambda_i\right)^\dagger\left(\bar e_L C_{\rm SRR}^{(6)} \nu + \bar e_R C_{\rm SRL}^{(6)} \nu\right)_{}^\dagger\right\}\,,\nn\\
s-ip_{\rm } &=& \left(s+ip \right)^\dagger\,,\nn\\
l^\mu_{ } &=& \frac{2 G_F}{\sqrt{2}}\lambda_i \left(\,\bar e_R \gamma^\mu C_{\rm VLR}^{(6)} \nu+\,\bar e_L \gamma^\mu C_{\rm VLL}^{(6)} \nu  \right)_{}+{\rm h.c.}\,,\nn\\
r^\mu_{} &=& \frac{2 G_F}{\sqrt{2}}\lambda_i \left(\,\bar e_R \gamma^\mu C_{\rm VRR}^{(6)} \nu+\,\bar e_L \gamma^\mu C_{\rm VRL}^{(6)} \nu  \right)_{}+{\rm h.c.}\,,\nn\\
t^{\mu\nu}_{L}&=&\frac{2 G_F}{\sqrt{2}}\bigg\{\lambda_i\,\bar e_R \sigma^{\mu\nu}C^{(6)}_{ \textrm{TLL}} \, \nu+\left(\lambda_i\right)^\dagger\left(\,\bar e_L \sigma^{\mu\nu}C^{(6)}_{ \textrm{TRR}} \, \nu \right)_{}^\dagger\bigg\}\,,\nn\\
t^{\mu\nu}_R&=&\left(t^{\mu\nu}_L\right)^\dagger\,,
\eea
with  $i= d, s$ denoting  a down quark or a strange quark in the currents. The matrices $\lambda_i$ are given by 
\begin{equation}
\lambda_d=\bma
0&1&0\\
0&0&0\\
0&0&0
\ema\,,
\qquad \qquad\lambda_s=\bma
0&0&1\\
0&0&0\\
0&0&0
\ema\,.
\end{equation}

The leading-order chiral Lagrangian consists of the Lorentz- and chiral-invariant terms with the lowest number of derivatives
\be
\mathcal L_{\mathrm{Meson}} = \frac{F_0^2}{4 } \mathrm{Tr}\left[(D_\mu U)^\dagger (D^\mu U)\right]+ \frac{F_0^2}{4} \mathrm{Tr}\left[U^\dagger \chi + U \chi^\dagger\right]\,,
\ee
where $D_\mu U = \partial_\mu U - i l_\mu U + i U r_\mu\,$, $\chi = 2 B (M + s -ip)\,,$  $F_0 =92.1$ MeV \cite{Rosner:2015wva}, and $U$ is 
\be
U(x) = \mathrm{exp}\left(\frac{i\sqrt{2}\Pi(x)}{\ F_0}\right)\,,\qquad \Pi(x) = \bma \frac{\pi^0}{\sqrt{2}}+\frac{\eta}{\sqrt{6}} &\pi^+ & K^+\\\pi^-&-\frac{\pi^0}{\sqrt{2}}+\frac{\eta}{\sqrt{6}} & K^0 \\
K^- & \bar{K}^0 &-\sqrt{\frac{2}{3}}\eta \ema \,.
\ee
The contribution from tensor sources first appears at $\mathcal{O} (p^4)$ and  it can not contribute to $K^-\rightarrow\pi^+l^-l^-$ at tree level. Hence we ignore the tensor sources in this part. While for the remaining sources, we  expand $U(x)$ to the leading order and the interactions relevant with the decay $\nu_j \rightarrow \pi^+ +e^-_i$  are  
\begin{equation}
\begin{aligned}
\mathcal L_\pi &=G_F F_0 \partial^\mu \pi^- \left\{ \bar{e}_{R, i} \gamma^\mu \nu_{j} (C_{\rm VRR}^{(6)}-C_{\rm VLR}^{(6)} )_{ udij}+\bar{e}_{L,i} \gamma^\mu \nu_j (C_{\rm VRL}^{(6)}-C_{\rm VLL}^{(6)} )_{ udij}  \right\} \\
&
-i F_0   B  G_F \pi^-\left\{ \bar{e}_{L,i}\nu_j (C_{\rm SLR}^{(6)}-C_{\rm SRR}^{(6)})_{udij}+\bar{e}_{R,i}\nu_j (C_{\rm SLL}^{(6)}-C_{\rm SRL}^{(6)})_{udij} \right\}\,,
\end{aligned}
\end{equation}
from which we  replace the index $d$ with $s$ and $\pi^-$ with $K^-$ to  get the  operators relevant with  the decay $K^-\rightarrow e_i^- + \nu_j$. 
 By contracting the neutrino in mass basis,  we  connect the operators containing a $\pi^-$ with those containing a $K^-$  and  get   the amplitude. For the process $K^-(k)\rightarrow\pi^+(p)l^-_1(p_1)l^-_2(p_2)$ where $l_{1,2}$ is an electron or a muon, there are two types of Feynman diagrams. They are different in the positions of the outgoing charged leptons. For the type where $l_1$ and $K^-$ share the same vertex, we get the amplitude
\begin{equation}
\begin{aligned}\label{eq:longamp}
\mathcal M_1 =-\frac{i F^2_0 G^2_F}{q^2-m_i^2}\Big\{
& m_iB^2\left(C^{(6)}_{\textrm{SLR}}-C^{(6)}_{\textrm{SRR}}\right)_{usl_1i}
\left(C^{(6)}_{\textrm{SLR}}-C^{(6)}_{\textrm{SRR}}\right)_{udl_2i}\bar{u}(p_1) P_R u^c(p_2)\\
&+ m_iB\left(C^{(6)}_{\textrm{VLR}}-C^{(6)}_{\textrm{VRR}}\right)_{usl_1i}
\left(C^{(6)}_{\textrm{SRR}}-C^{(6)}_{\textrm{SLR}}\right)_{udl_2i}\bar{u}(p_1) \slashed k P_R u^c(p_2)\\ 
&+ m_iB\left(C^{(6)}_{\textrm{SLL}}-C^{(6)}_{\textrm{SRL}}\right)_{usl_1i}
\left(C^{(6)}_{\textrm{VRL}}-C^{(6)}_{\textrm{VLL}}\right)_{udl_2i}\bar{u}(p_1) \slashed p P_R u^c(p_2)\\ 
&- B^2\left(C^{(6)}_{\textrm{SLL}}-C^{(6)}_{\textrm{SRL}}\right)_{usl_1i}
\left(C^{(6)}_{\textrm{SLR}}-C^{(6)}_{\textrm{SRR}}\right)_{udl_2i}\bar{u}(p_1) \slashed q P_R  u^c(p_2)\\ 
&+ m_i\left(C^{(6)}_{\textrm{VLL}}-C^{(6)}_{\textrm{VRL}}\right)_{usl_1i}
\left(C^{(6)}_{\textrm{VLL}}-C^{(6)}_{\textrm{VRL}}\right)_{udl_2i}\bar{u}(p_1) \slashed k \slashed p P_R u^c(p_2)\\ 
&- B\left(C^{(6)}_{\textrm{VLL}}-C^{(6)}_{\textrm{VRL}}\right)_{usl_1i}
\left(C^{(6)}_{\textrm{SRR}}-C^{(6)}_{\textrm{SLR}}\right)_{udl_2i}\bar{u}(p_1) \slashed k \slashed q P_R u^c(p_2)\\ 
&- B\left(C^{(6)}_{\textrm{SRR}}-C^{(6)}_{\textrm{SLR}}\right)_{usl_1i}
\left(C^{(6)}_{\textrm{VLL}}-C^{(6)}_{\textrm{VRL}}\right)_{udl_2i}\bar{u}(p_1) \slashed q \slashed p P_R u^c(p_2)\\ 
&- \left(C^{(6)}_{\textrm{VLR}}-C^{(6)}_{\textrm{VRR}}\right)_{usl_1i}
\left(C^{(6)}_{\textrm{VLL}}-C^{(6)}_{\textrm{VRL}}\right)_{udl_2i}\bar{u}(p_1) \slashed k \slashed q \slashed p P_R u^c(p_2)
\,\\
&+\rm{(terms\, by \, flipping\, the\, chirality\, of\, leptons)}\Big\}\,,
\end{aligned}
\end{equation}
where $q=k-p_1$,  $u(p_{1,2})$ denotes a spinor with momentum $p_{1,2}$, $m_i$ is the Majorana neutrino mass below GeV scale and a summation over $i$ is implied. In the second type of Feynman diagrams  $l_2$ and $K^-$ share the same vertex. To get its amplitude $\mathcal M_2$, we only need the replacement $p_1 \leftrightarrow p_2$ and add one minus sign in $\mathcal M_1$.  The total amplitude is
\begin{equation}
\mathcal M_L = \mathcal{M}_1 -\mathcal{M}_1\Big|_{p_1\leftrightarrow p_2} \,.
\end{equation}
The amplitude  for the term $C^{(6)}_{\textrm{VLL},usl_1 i}\times C^{(6)}_{\textrm{VLL},udl_2i}$ is four times that of Ref. \cite{Liao:2020roy} and  the h.c. terms    of $l_\mu$ and $r_\mu$ were missed.  

In principle this is not the whole story. The exchange of virtual sterile neutrinos with small masses but hard momenta (larger than $\Lambda_\chi$) leads to hadronic LNV operators without neutrinos. This is very similar to the exchange of virtual hard photons gives rise to the mass splitting between charged and neutral pions. This so-called hard-neutrino exchange plays an important role in nuclear $0\nu\beta\beta$ and has been the focus of a lot of recent work \cite{Cirigliano:2018hja,Cirigliano:2019vdj,Cirigliano:2020dmx,Richardson:2021xiu,Cirigliano:2021qko,Wirth:2021pij,Jokiniemi:2021qqv}. These contributions can readily be incorporated for the LNV kaon decays  as well (see e.g. Ref.~\cite{Dekens:2020ttz} for operators involving pions). As we will discuss below, the contributions from the exchange of off-shell neutrinos are not sufficiently large to give meaningful constraints. We therefore do not construct the corresponding operators. However, their corrections should be included when the sterile neutrino mass is outside the resonance region.

\subsection{The case of heavy sterile neutrinos}
 The chiral Lagrangian induced by the dimension-9 operators in Eq.~\eqref{eq:Lag9} has been discussed in Refs.\cite{Liao:2019gex}. The most relevant hadronic interactions involve one pion, one kaon  and  two charged leptons. The mesonic chiral Lagrangian is
 \begin{equation}
 \begin{aligned} \mathcal{L_{S}}=&\frac{F^4_0}{4 v^5}\Bigg[ \frac{5}{3}g_1^{\pi K}C^{(9)}_{1\,L/R}L^\mu_{21}L_{31\,\mu}+(g_2^{\pi K}C^{(9)}_{2\,L/R}+g_3^{\pi K} C^{(9)}_{3\,L/R})U_{21}U_{31}\\
 &+(g_4^{\pi K}C^{(9)}_{4\,L/R}+g_5^{\pi K} C^{(9)}_{5\,L/R})U_{21}U^\dagger_{31}\Bigg]\bar{e}_{L/R} C\bar{e}^T_{L/R}\\
 &+\frac{F^4_0}{4 v^5}\bar{e}\gamma_\mu\gamma_5 C\bar{e}^T\Bigg[(g_6^{\pi K}C^{(9)}_{6,\rm usud}+g_7^{\pi K} C^{(9)}_{7,\rm usud})L^\mu_{31}U^\dagger_{21}+(g_6^{\pi K\prime}C^{(9)}_{6,\rm udus}+g_7^{\pi K\prime} C^{(9)}_{7,\rm udus})L^\mu_{21}U^\dagger_{31}\\
  &+g^{\pi K}_{8/9}( C^{(9)}_{8/9,\rm usud}+ C^{(9)}_{8/9,\rm udus})(L^\mu_{31}U_{21}+L^\mu_{21}U_{31})\\
  &+g^{\pi K\prime}_{8/9}( C^{(9)}_{8/9,\rm usud}- C^{(9)}_{8/9,\rm udus})(L^\mu_{31}U_{21}-L^\mu_{21}U_{31})\Bigg]\\
&+(C^{(9)}_i\rightarrow C^{(9)\prime}_i)\,,
 \end{aligned}
 \end{equation}
 where $L_\mu=i UD_\mu U^\dagger$ and the parity invariance of QCD implies that the hadronic operators induced by $O_i$ are the same as those induced by $O_i^\prime$  and they share the same LECs.  Then we expand the $U$ matrix and get operators involving two mesons and two leptons \cite{Liao:2019gex}
  \begin{equation}
 \begin{aligned} \mathcal{L_{S}}=&\frac{1}{v^5}K^-\pi^- \left [c_1 \bar{e}_LC\bar{e}^T_L +c_2 \bar{e}_RC\bar{e}^T_R\right]+\frac{1}{v^5}\left[c_3 \partial^\mu K^-\pi^-+c_4\partial^\mu \pi^- K^- \right]\bar{e}\gamma_\mu \gamma_5 C\bar{e}^T\\
 &+\frac{1}{v^5}\partial^\mu K^- \partial_\mu \pi^- \left[c_5 \bar{e}CP_L\bar{e}^T +c_6 \bar{e}CP_R\bar{e}^T  \right]\,,
 \end{aligned}
 \end{equation}
 where the parameters $c_i$ are 
 \begin{equation}
 \begin{aligned}
 c_1\, =\, &-\frac{1}{2} F^2_0[g^{\pi K}_{2}(C^{(9)}_{2 \,L}+C^{(9)\prime}_{2\,L})+ g^{\pi K}_{3}(C^{(9)}_{3\,L}+C^{(9)\prime}_{3\,L})\\
 &- g^{\pi K}_{4}(C^{(9)}_{4\,L}+C^{(9)\prime}_{4\,L})  - g^{\pi K}_{5}(C^{(9)}_{5\,L}+C^{(9)\prime}_{5\,L}) ]\,,\\
 c_2\,=\,&c_1\Big|_{L\rightarrow R}\,,\\
 c_3\,=\,&-\frac{i}{2}F^2_0 [g^{\pi K}_{6} (C^{(9)}_{6,usud}+C^{{(9)}\prime}_{6, usud})+g^{\pi K}_{7} (C^{(9)}_{7,usud}+C^{(9)\prime}_{7, usud})\\
 &-g^{\pi K}_{8/9} (C^{(9)}_{8/9,usud}+C^{(9)\prime}_{8/9, usud}+C^{(9)}_{8/9,udus}+C^{(9)\prime}_{8/9, udus})
 \\
 & -g^{\pi K\prime}_{8/9} (C^{(9)}_{8/9,usud}+C^{(9)\prime}_{8/9, usud}-C^{(9)}_{8/9,udus}-C^{(9)\prime}_{8/9, udus})
 ]\,,\\
 c_4\,=\,&-\frac{i}{2}F^2_0 [g^{\pi K\prime}_{6} (C^{(9)}_{6,udus}+C^{{(9)}\prime}_{6, udus})+g^{\pi K\prime}_{7} (C^{(9)}_{7,udus}+C^{(9)\prime}_{7, udus})\\
&-g^{\pi K}_{8/9} (C^{(9)}_{8/9,usud}+C^{(9)\prime}_{8/9, usud}+C^{(9)}_{8/9,udus}+C^{(9)\prime}_{8/9, udus})
\\
& +g^{\pi K\prime}_{8/9} (C^{(9)}_{8/9,usud}+C^{(9)\prime}_{8/9, usud}-C^{(9)}_{8/9,udus}-C^{(9)\prime}_{8/9, udus})
]\,,\\
 c_5\,=\,&\frac{5}{6} F^2_0 g^{\pi K}_{1} (C^{(9)}_{1\,L}+C^{(9)\prime}_{1\,L})\,,\\
 c_6\,=\,&c_5\Big|_{L\rightarrow R}\,.
 \end{aligned}
 \end{equation}
The LECs, $g^{\pi K}_i$, can be estimated by using naive dimensional analysis (NDA) 
\begin{equation}
g_1^{\pi K}=\mathcal O(1)\,,\qquad g^{\pi K}_{2,3,4,5}=\mathcal O(\Lambda^2_\chi)\,,\qquad g^{\pi K(\prime)}_{6,7,8,9}= \mathcal O(\Lambda_\chi)\,.
\end{equation}
 We can also  relate $g^{\pi K}_i$ with the LECs appearing in $K^0\rightarrow \bar{K}^0$ \cite{Cirigliano:2017ymo}, $K^\pm\rightarrow\pi^\pm\pi^0$ \cite{Savage:1998yh} and  $\pi^-\rightarrow\pi^+$ \cite{Cirigliano:2017ymo},  some of which have been computed by several lattice QCD groups \cite{Carrasco:2015pra,ETM:2012vvy,SWME:2015oos,Boyle:2012qb,Garron:2016mva,Blum:2015ywa,Blum:2012uk}. 
The tree-level amplitude for $K^-(k)\rightarrow\pi^+(p)l^-(p_1)l^-(p_2)$  can be read off directly 
\bea\label{short}
\mathcal M_S &=& - i \frac{1}{v^5}\Bigg[2 c_1 \bar{u}(p_1)P_R u^c(p_2)+2 c_2 \bar{u}(p_1)P_L u^c(p_2)-2 i c_3\bar{u}(p_1) \slashed k \gamma_5  u^c(p_2)\nn\\
&+&2 i c_4\bar{u}(p_1) \slashed p \gamma_5  u^c(p_2)
+2 k\cdot p [c_5 \bar{u}(p_1)P_R u^c(p_2)+ c_6 \bar{u}(p_1)P_Lu^c(p_2) ]
\Bigg]\,.
\eea

\section{Phase space integral}\label{sec:int}
The momentum products in the amplitude square $|\mathcal M|^2$ of the decay  $K^-(k)\rightarrow\pi^+(p)l^-_1(p_1)l^-_2(p_2)$ have two independent terms, $(k-p_1)^2=q^2$ and $k\cdot p_2$, and all the other products can be expressed in terms of these two products and particle masses. To simplify the integral further, we write \cite{Liao:2020roy}
\begin{equation}
(k-p_1)^2=a\,,\qquad k\cdot p_2=\frac{1}{4a}(m_K^2+a-m_{l_1}^2)(a+m_{l_2}^2-m_\pi^2)+\frac{1}{\sqrt{a}}m_K |\vec{q}||\vec{p_2}|\cos\theta\,,
\end{equation}
where $|\vec{q}|=\frac{\lambda^{\frac{1}{2}}(m_K,\sqrt{a},m_{l_1})}{2m_K}$, $|\vec{p_2}|=\frac{\lambda^{\frac{1}{2}}(m_\pi,\sqrt{a},m_{l_2})}{2\sqrt{a}}$ with $\lambda(a,b,c)=a^4+b^4+c^4-2a^2b^2-2a^2c^2-2b^2c^2$ and $m_{l_i}$ is the mass of lepton $l_i$.
The decay rate becomes
\begin{equation}\label{eq:int}
\Gamma =(2-\delta_{l_1l_2})\frac{1}{2!} \frac{1}{2m^2_K}\frac{1}{64\pi^3}\int da\int d\!\cos\theta\, |\vec{q}|\frac{|\vec{p_2}|}{\sqrt{a}}|\mathcal M|^2(a,\,k\cdot p_2)\,,
\end{equation}
 and the integration domains are given by
\begin{equation}
\begin{aligned}
a \in& [(m_{l_2}+m_\pi)^2\,,\,(m_K-m_{l_1})^2]\,,\qquad
\cos\theta \in& [-1\,,\,1]\,.
\end{aligned}
\end{equation}

If the mass of neutrino is in the range $ [m_{l_{1,2}}+m_\pi\,,\,m_K-m_{l_{2,1}}]$, the exchanged neutrino can go on shell. 
Near the pole, we modify the propagator
 \begin{equation}
 \frac{1}{q^2-m_i^2+i\epsilon}\longrightarrow \frac{1}{q^2-m_i^2+im_i \Gamma_{i}}\,,
 \end{equation}
 where $\Gamma_{i}$ is the total decay width of $\nu_i$ in the mass basis. When the mass of sterile neutrino is much larger than its decay rate,  we  use the narrow width approximation 
 \begin{equation}
 \frac{1}{(q^2-m^2_i)^2+m_i^2 \Gamma_{i}^2}\longrightarrow \frac{\pi}{m_i\Gamma_{i}}\delta(q^2-m^2_i)\,,
 \end{equation}
to simplify the phase space. The resulting $m_i/\Gamma_i$ enhancement is typically large enough that other contributions can be neglected. 

\section{Phenomenology}\label{sec:pheno}
The EFT approach to the long-distance contributions without a sterile neutrino  and the short-distance contributions has been discussed in Refs. \cite{Liao:2020roy, Liao:2019gex}, and the current experiments  can only set a loose bound on the BSM energy scale $\Lambda$ with $\Lambda > \mathcal O(10)$ GeV. This scale is too low for the SM-EFT approach to be valid. However, this is not the case when we work in the framework of $\nu$SMEFT. In this section  we first show the effect of  the sterile neutrino on the short-distance contribution and  then discuss two scenarios for the long-distance contribution and  instruct the resonance enhancement. Finally we give constraints on the energy scale of operators in table \ref{tab:O6R}.

\subsection{Short-distance contribution  }
Ref. \cite{Liao:2019gex} considered \textoverline{ dim-7} LNV operators in SMEFT and matched them onto dim-9 operators ($C^{(9)}_{5\,L}$, $C^{(9)\prime}_{5\,L}$ and $C^{(9)}_{1\,L}$), and the resulting  $C^{(9)}_{i}$ are proportional to $\frac{v^3}{\Lambda^3}$  given that the WCs of \textoverline{dim-n} operators are proportional to $\frac{1}{\Lambda^n}$. Using the current experimental limit, they obtained a relatively weak bound on $\Lambda>\mathcal O$(10) GeV. 

In the presence of a sterile neutrino with  mass $m_\nu>\Lambda_\chi$, where the same $C^{(9)}_{i}$ are induced, the bound on $ \Lambda$ can be slightly improved. 
For instance, let us consider a dim-9 operator, $C^{(9)}_{5\,R}$, which could receive a contribution from two dim-6 operators  $ C^{(6)}_{\text{VRR,us}}\times  C^{(6)}_{\text{VLR,ud}}$, or equivalently from two \textoverline{dim-6} operators  $\mathcal{O}^{(6)}_{du\nu e}\times \mathcal{O}^{(6)}_{H\nu e}$.    $C^{(9)}_{5\,R}$ is thus proportional to $v^5/(m_\nu \Lambda^4)$. Due to the  enhancement by $v/m_\nu$ it is possible to get a more stringent bound.For instance, for $m_\nu =$ 1 GeV,  we get a better lower limit on $\Lambda$ with $\Lambda> \mathcal{O}$(100) GeV based on current experimental limits, which improves the result in Ref. \cite{Liao:2019gex} by one order. Nevertheless, it is clear that for sterile neutrinos with masses above a GeV or so, the resulting limits are rather weak and it is unclear whether the use of the SMEFT or $\nu$SMEFT frameworks are justified.

\subsection{Long-distance contributions and resonances}
If  a sterile neutrino exists with a mass inside the resonance range $[m_l+m_\pi,m_K-m_l]$, the decay rate is significantly enhanced \cite{Dib:2000wm} and we can get  much stronger constraints on $\Lambda$ and neutrino mixing angles. In this subsection we ignore  sterile neutrinos with mass outside the resonance range and  discuss two scenarios, the minimal scenario and the  leptoquark scenario, and show their effects on the decay rates of   $K^-(k)\rightarrow\pi^+(p)l^-(p_1)l^-(p_2)$. 
\subsubsection{The minimal scenario}
In the minimal scenario, we add a sterile neutrino $\nu_R$ with  mass $m_\nu$ in the resonance region and it can only interact with the SM particles via the mixing with active neutrinos. We get the Lagrangian by setting  the WCs of operators from tables~\ref{tab:O6L}-\ref{tab:O7R} to zero and writing the active neutrino $\nu_\alpha$ in the weak interaction in terms of the neutrino mass eigenstates $\nu_i$
\begin{equation}
\nu_\alpha =U_{\alpha i}\nu_i\,,
\end{equation}    
where $\alpha =e, \mu$ and $i=1,2,3,4$.
  We also assume the sterile neutrino   mixes only with the electron  neutrino $\nu_e$  in $K^-\rightarrow \pi^+e^-e^- $ or with the muon neutrino $\nu_\mu$ in $K^-\rightarrow \pi^+\mu^-\mu^- $.  Due to the small mixing angles between the sterile- and active- neutrinos, $\nu_R$ is approximately equivalent to $\nu_4$.

\begin{figure}[t]
 	\begin{center}
 	\includegraphics[scale=0.65]{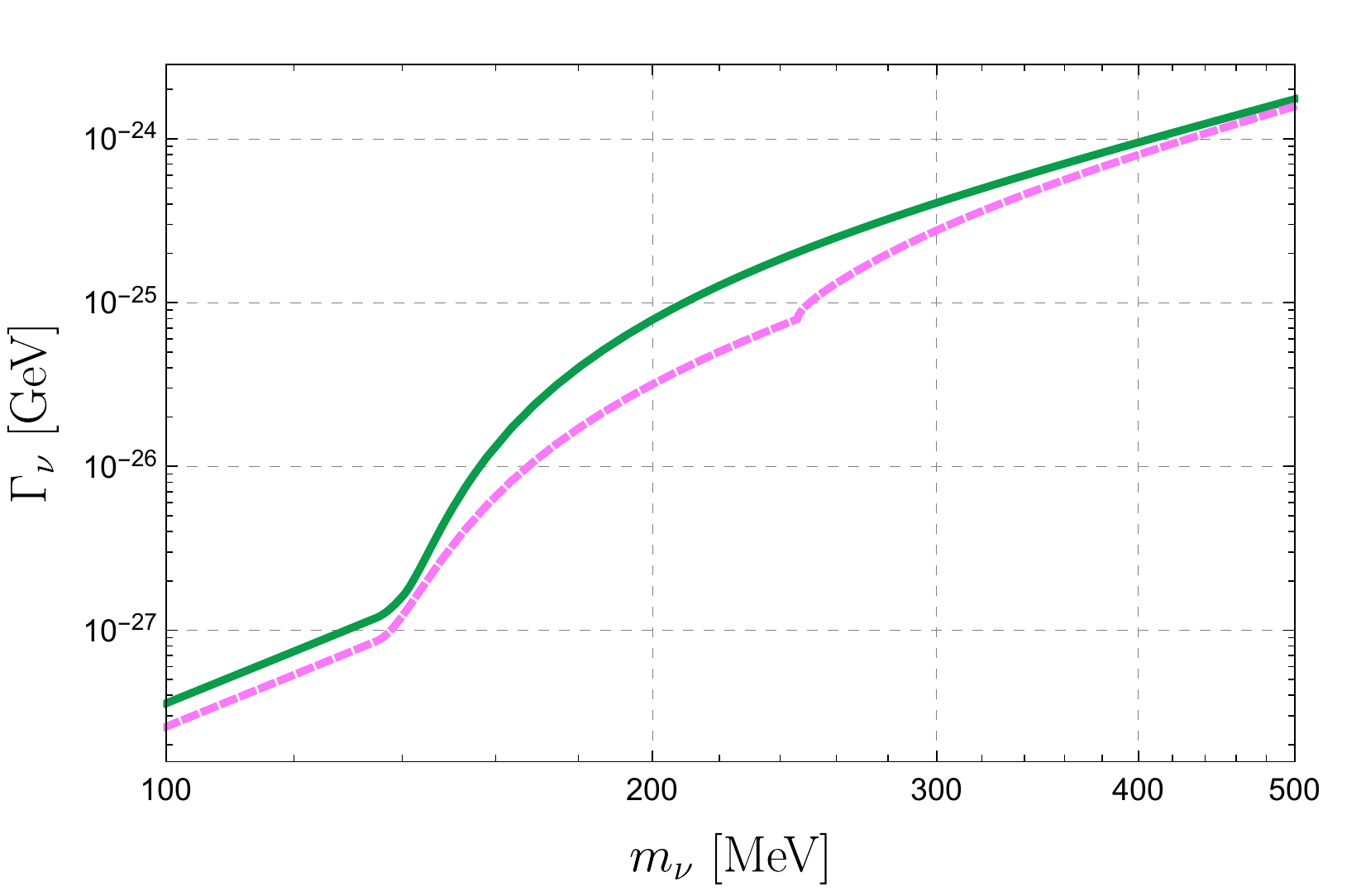}
 	\caption{ Decay rates of the sterile neutrino in the minimal scenario for cases  $U_{e4}\ne 0$ (green solid) and $U_{\mu 4}\ne 0$ (pink dashed ). The kinks at $m_\nu \approx$ 140 and 240 MeV are due to threshold of decay channels $N\rightarrow \pi^0+ \nu_e/\nu_\mu $ and $\pi^\pm+e^\mp/\mu^\mp$.    }\label{fig:gammaSM}
 \end{center}
 \end{figure}

\begin{figure}[t]
	\includegraphics[scale=0.36]{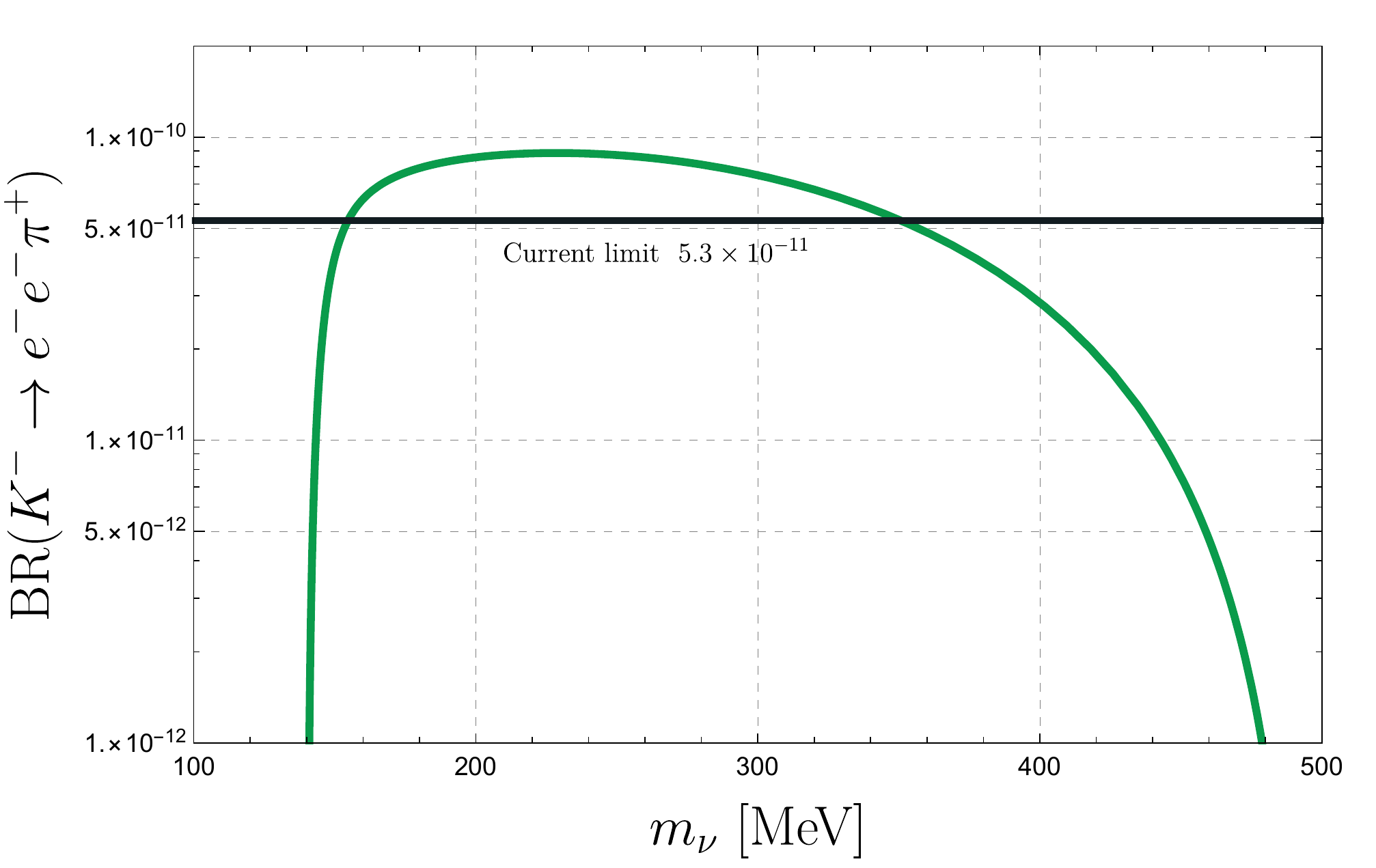}
		\includegraphics[scale=0.36]{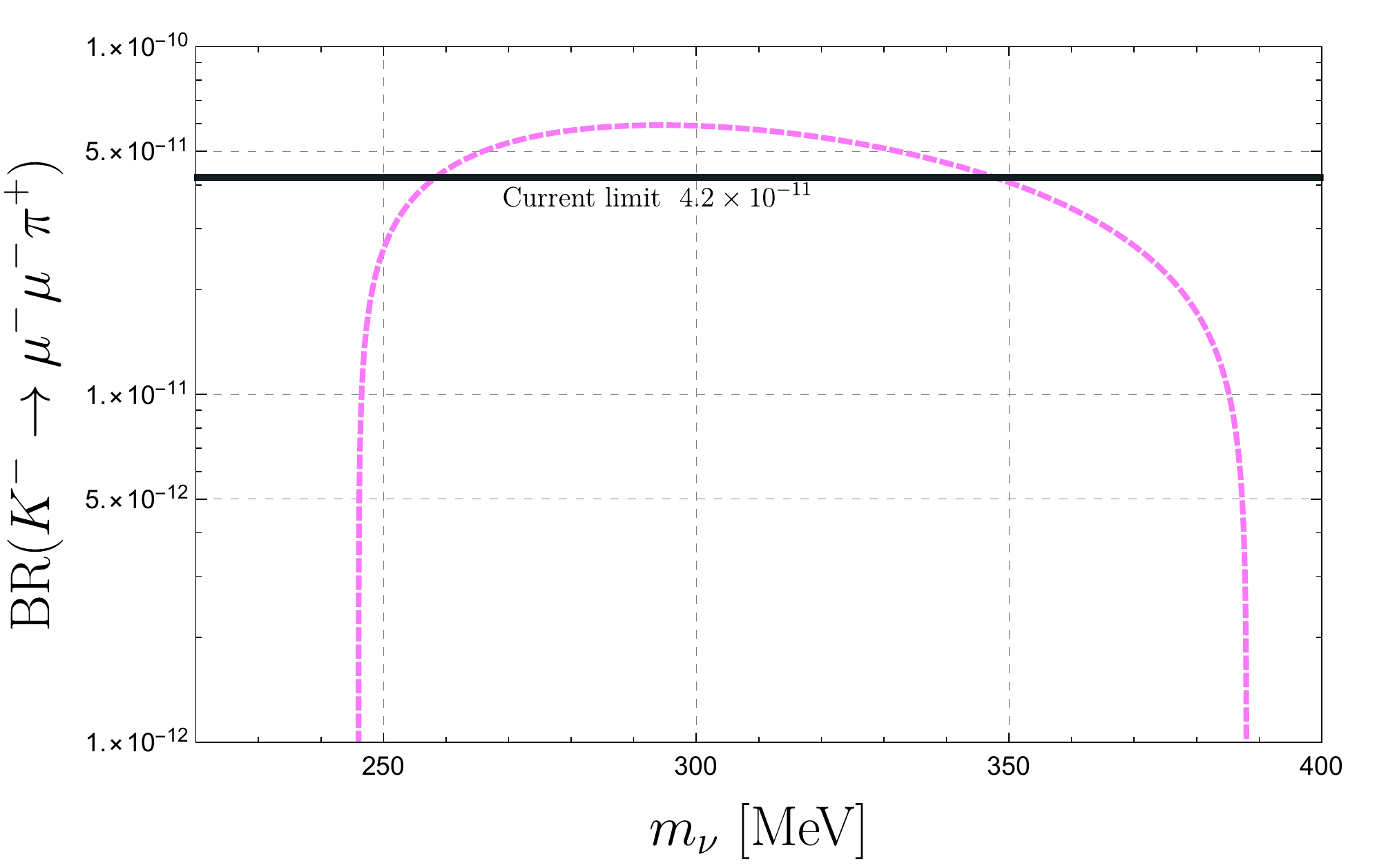}
	\caption{ Branching ratios of $K^-\rightarrow\pi^+e^-e^-$ (left panel) and $K^-\rightarrow\pi^+\mu^-\mu^-$ (right panel ) as functions of the sterile neutrino mass $m_\nu$ in the minimal scenario. }\label{fig:brSM}
\end{figure}

   The possible decay modes of the sterile neutrino are discussed in  App. \ref{app:decay}. We show the decay rates of the sterile neutrino in Fig.~\ref{fig:gammaSM}. In Fig.~\ref{fig:brSM} we plot the branching ratios of kaons as a function of $m_\nu$ for the case of final-state electrons (left panel) and muons (right panel).
  When calculating the decay rates, the mixing angles $|U_{e4}|$ and $|U_{\mu4}|$ are set to the see-saw prediction $\sqrt{0.05\,\,\rm{eV}/m_\nu}$  with $m_\nu$ the mass of the sterile neutrino. 
 It is clear that $m_\nu/\Gamma_N \gg 1$ and it is safe to use the narrow-width approximation.  The two branching ratios are slightly above the current limit  around 300 MeV. Hence either there is no such a sterile neutrino with a Majorana mass  around 300 MeV, or
  the mixing angle $|U_{l4}|$ ($l=e\,,\mu$) is smaller than the see-saw relation.  
  
\begin{figure}[t]
	\includegraphics[scale=0.5]{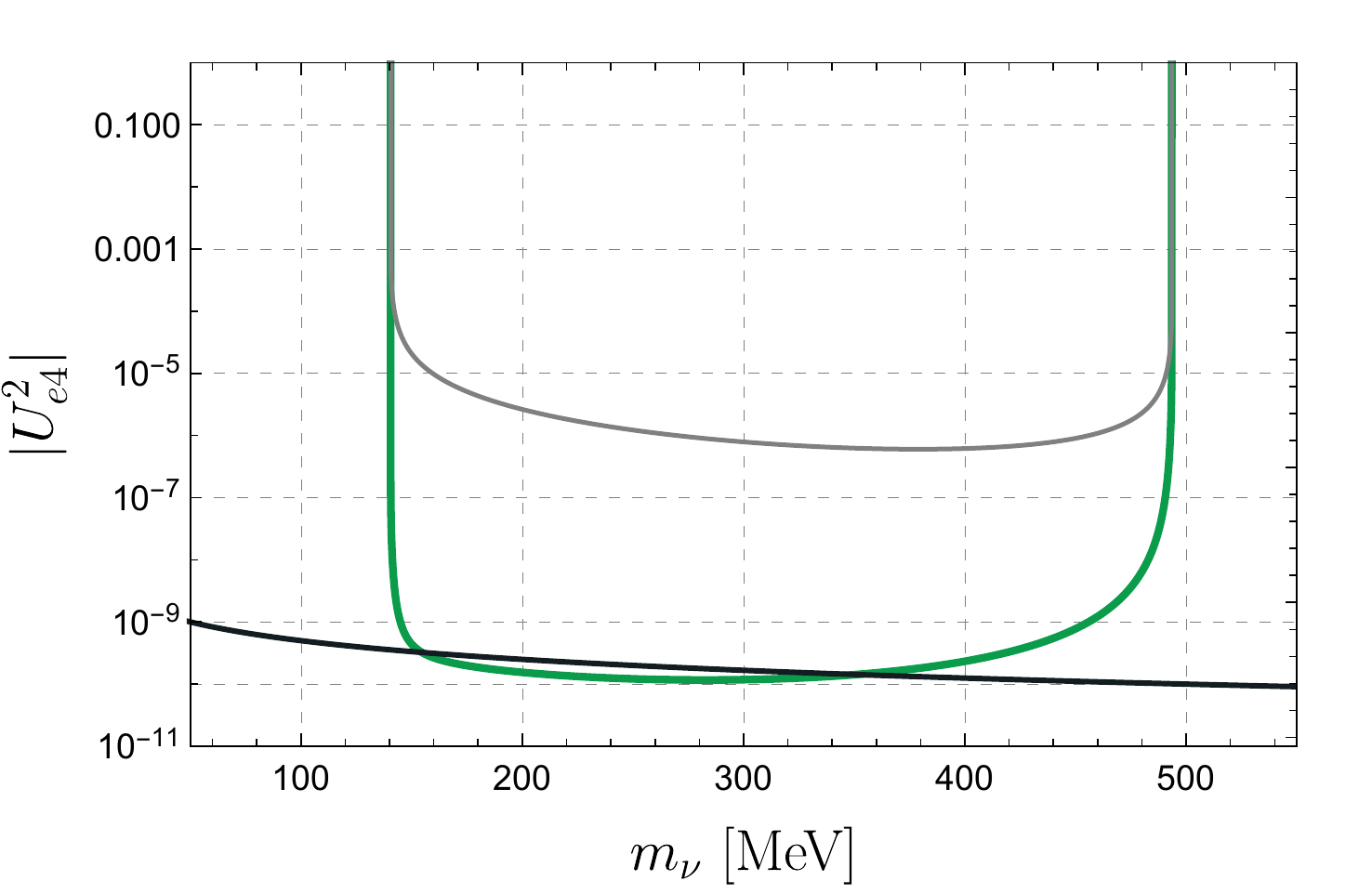}
	\includegraphics[scale=0.5]{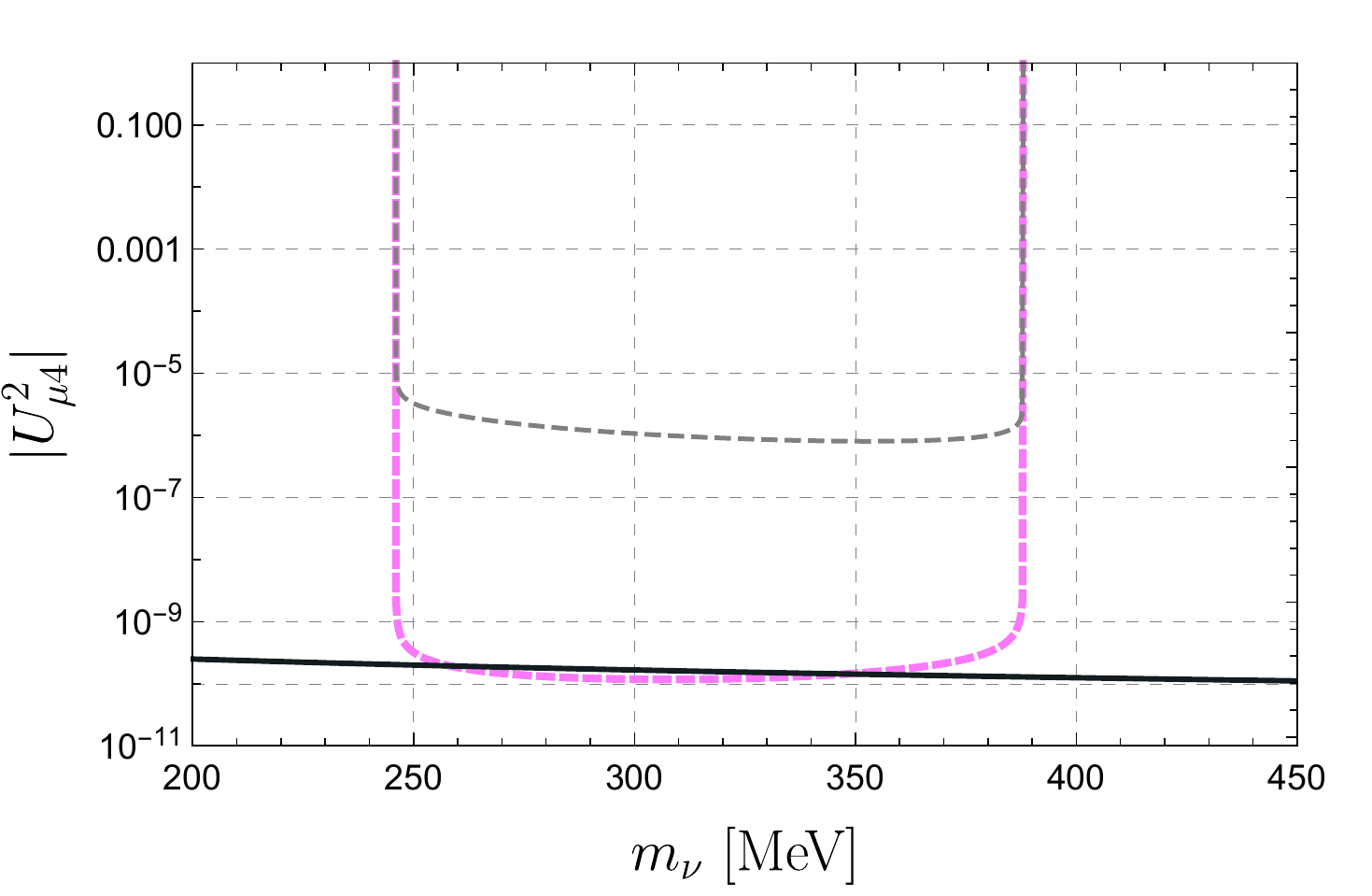}
	\caption{ The excluded parameter space above the curves for $|U_{e4}|^2$ (left panel) and $|U_{\mu4}|^2$ (right panel) from the limits on  branching ratios of $K^-\rightarrow\pi^+l^-l^-$.  The two black lines correspond to the type-I seesaw  relation  $\{|U_{e4}|^2,\,|U_{\mu4}|^2\}=0.05 \,\,\rm eV/m_\nu$. The gray  (dashed) lines are the modified limits when including the finite detector size effect. }\label{fig:limitSM}
\end{figure}

  In Fig.~\ref{fig:limitSM} we show  the limits on $|U_{e4}|^2$ and    $|U_{\mu 4}|^2$ as functions of $m_\nu$. The limits are quite close to the black curve indicating the type-I seesaw relation. The constraints on $|U_{e 4}|^2$ and $|U_{\mu 4}|^2$ could reach  $\mathcal O(10^{-10})$.    The limits on the mixing angles detoriate quickly near the boundaries of resonance regions due  to the phase space suppression.

\subsubsection{The leptoquark scenario}
In this section we neglect the interactions of the minimal scenario and  the SM is extended only by interactions with leptoquarks (LQs), which can convert quarks to leptons and vice versa. Ref. \cite{Dorsner:2016wpm} summarized all the possible representations of LQs and we focus on a scalar LQ$: \tilde R\left({\bf 3},~{\bf 2},~1/6\right)$. Its interactions with quarks and leptons are given by
\begin{align}\label{eq:LQlag}
{\cal L}_{\rm LQ}=-{y}^{RL}_{ij}\bar{d}_{Ri}\tilde R^a\epsilon^{ab}L_{Lj}^{b}+y^{\overline {LR}}_{kl}\bar{Q}^{a}_{Lk}\tilde R^a\nu_{Rl} +{\rm h.c.}\,,
\end{align}
where  $i,j, k, l$ and $a,b$ are flavor and $SU(2)$ indices, respectively. After integrating the LQ we get the following \textoverline{dim-6} operator
\begin{align}
{\cal L}^{(\bar{6})}_{\nu_R}=C^{({6})}_{LdQ\nu,  ijkl}\left(\bar{L}^a_id_j \right)\epsilon^{ab}\left(\bar{Q}^b_k\nu_{Rl} \right)+{\rm h.c.}\,,
\end{align}
where
\begin{align}
C^{(\bar{6})}_{LdQ\nu, ijkl}=\frac{1}{m^2_{\rm LQ}}y^{\overline{LR}}_{kl}y^{RL*}_{ji}\,,
\end{align}
and $m_{\rm LQ}$ is the mass of the LQ. Below the electroweak scale, four operators are induced
\begin{align}
{\cal L }^{(6)}_{\Delta L=0}=\frac{2G_F}{\sqrt{2}}\bigg[\bar{c}_{\textrm{SR}, ijkl}^{(6)}~\bar{u}_{L,i}d_{R, j}\bar{e}_{L, k}\nu_{R,l}+\bar{c}^{(6)}_{ \textrm{T}, ijkl}~\bar{u}_{L,i} \sigma^{\mu\nu}d_{R, j}\bar{e}_{L, k}\sigma^{\mu\nu}\nu_{R,l}\,\nn\\
+\bar{c}_{\textrm{NSR}, ijkl}^{(6)}~\bar{d}_{L,i}d_{R, j}\bar{\nu}_{L, k}\nu_{R,l}+\bar{c}^{(6)}_{ \textrm{NT}, ijkl}~\bar{d}_{L,i} \sigma^{\mu\nu}d_{R, j}\bar{\nu}_{L, k}\sigma^{\mu\nu}\nu_{R,l}
\bigg]+{\rm h.c.}\,,
\end{align}
where the two neutral currents  contribute to the decay width of the sterile neutrino and thus affect  LNV decay process in the resonance region,  and   the coefficients satisfy
\begin{align}
\bar{c}^{(6)}_{\textrm{ SR}, ijkl}=4\bar{c}^{(6)}_{\textrm{T}, ijkl}=\frac{v^2}{2m^2_{\rm LQ}}y^{\overline{LR}}_{il}y^{RL*}_{jk}\,,\\
\bar{c}^{(6)}_{\textrm{ NSR}, ijkl}=4\bar{c}^{(6)}_{\textrm{NT}, ijkl}=\frac{v^2}{2m^2_{\rm LQ}}y^{\overline{LR}}_{ml}y^{RL*}_{jk} V^*_{mi}\,.
\end{align}
The matching to the operators in eq. (\ref{6final}) is 
\begin{align}
C^{(6)}_{\textrm{SRR},ijkl}=4C^{(6)}_{\textrm{TRR},ijkl}=\sum^n_{l=1} \bar{c}^{(6)}_{\textrm{ SR},ijkl}U^*_{3+l,i}\,,
\end{align}
where $n$ is the number of sterile neutrinos and here we consider only one sterile neutrino. Since we focus on the resonance region, the contributions from other light neutrinos can be safely ignored and the mixing angle $|U_{44}|\approx 1$.

\begin{figure}[t]
	\begin{center}
	\includegraphics[scale=0.85]{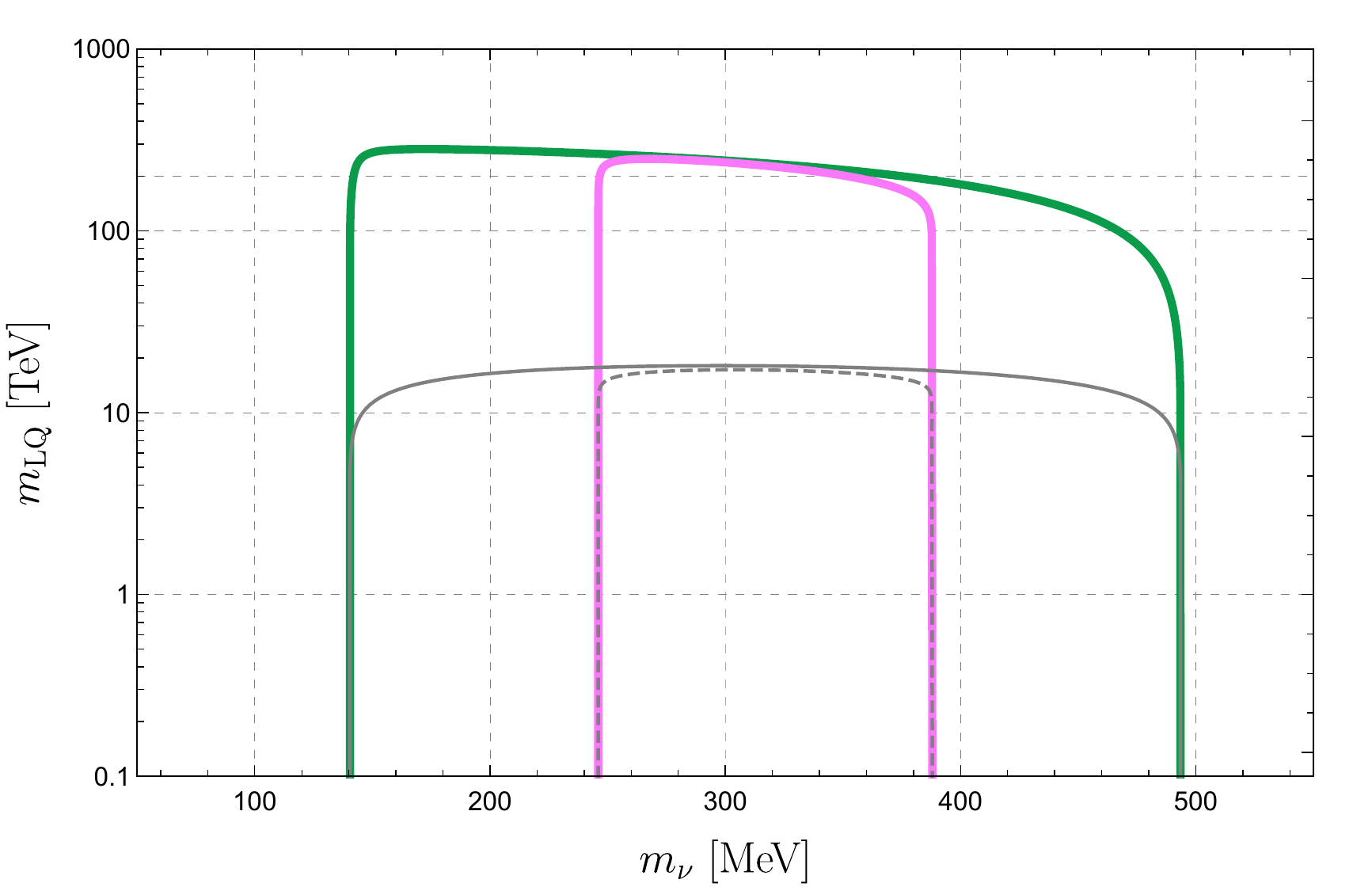}
	\caption{Limits on the leptoquark mass $m_{\text{LQ}}$ versus the sterile neutrino mass $m_\nu$ in the resonance region. The green curve is based on current limit for $K^-\rightarrow\pi^+e^-e^-$ and similarly the pink one for $K^-\rightarrow\pi^+\mu^-\mu^-$. The gray (dashed) lines include the finite detector size effect.  }\label{fig:limitlq}
\end{center}
\end{figure}

In order to induce the decay $K^-\rightarrow\pi^+ e^-e^-$, we set $y^{\overline{LR}}_{u1}y^{RL*}_{de}=y^{\overline{LR}}_{u1}y^{RL*}_{se}$ to one and all other indices configurations to zero. Similarly for $K^-\rightarrow\pi^+ \mu^-\mu^-$  we assume $y^{\overline{LR}}_{u1}y^{RL*}_{d\mu}=y^{\overline{LR}}_{u1}y^{RL*}_{s\mu}=1$  with all others being zero. Then the  decay rates  are a function of the leptoquark mass $m_{\text{LQ}}$  and the  neutrino mass $m_\nu$.
 One can check the decay rate of the sterile neutrino is  much smaller than its mass and thus the  narrow width approximation is still valid.
 In Fig. \ref{fig:limitlq}, we show the limits on $m_{\text{LQ}}$ by varying $m_\nu$. The regions below the two colorful curves are excluded,  and the green curve ($K^-\rightarrow\pi^+ e^-e^-$) reaches an energy scale around 300 TeV while the pink curve ($K^-\rightarrow\pi^+ \mu^-\mu^-$) could reach 250 TeV. Due to the same reason as that in the minimal scenario, $m_{\text{LQ}}$ approaches 0 near the resonance boundaries.

 \subsection{Limits on \textoverline{dim-6} operators with a sterile neutrino}
 In  principle by using the current limit on $K^-\rightarrow\pi^+l^-l^-$ in the resonance region we can make a limit plot for every operator  from tables \ref{tab:O6L}-\ref{tab:O7R}. The strongest limits arise from (part of) the operators in table \ref{tab:O6R},
 because  contributions from the operators in the other tables are suppressed by  either the small mixing angles $|U_{e/\mu4}|$ for those containing a left-handed neutrino  or $\Lambda$ for the \textoverline{dim-7} operators. Not all $\nu$SMEFT dimension-six operators contribute equally.  For instance, $\mathcal{O}^{(6)}_{L\nu H}$ has no direct effect on the LNV kaon decay and is  ignored here. $\mathcal{O}^{(6)}_{\nu W}$ can induce the decay $N\rightarrow\nu\gamma$, which is relatively fast and decreases the LNV kaon decay rates in the resonance region. $\mathcal{O}_{\nu W}^{(6)}$ is strictly constrained 
 because it generates neutrino dipole moments  at one-loop \cite{Butterworth:2019iff,Canas:2015yoa} and it is also ignored. 
We are then left with four operators  ($\mathcal O^{(9)}_{LdQ\nu}$ has been discussed in previous subsection) in table \ref{tab:O6R}.
These operators can easily be obtained  
in  models with $Z'$ bosons $(\mathcal O^{(9)}_{Qu\nu L}$, $\mathcal 
O^{(9)}_{L\nu Qd}$), and left-right symmetric models ($\mathcal O^{(9)}_{H\nu e}$). We refrain from introducing specific models and focus on giving the limits on the WCs of these  four operators directly.

To induce  the LNV kaon decay, we turn on a single operator with specific flavor configurations  at a time and ignore the minimal interactions. For  $\mathcal O^{(9)}_{Qu\nu L}$ and $\mathcal O^{(9)}_{H\nu e}$ we only need to turn on one flavor configuration to induce LNV kaon decay. We set $ C^{(9)}_{H\nu e,11} = \frac{1}{\Lambda^2}$  to induce $K^-\rightarrow\pi^+e^-e^-$  with all other indices configurations being zero. Then we can get  limits on $\Lambda$ as a function the sterile neutrino mass $m_\nu$. Because the left handed down-type quarks are not in mass eigenstates, $K^-\rightarrow\pi^+e^-e^-$ can also be realized via $C^{(9)}_{Qu\nu L,1111}=\frac{1}{\Lambda^2}$. 
The remaining two operators are special in the sense that we need to turn on two flavor configurations to induce LNV kaon decays. For convenience we assume $C^{(9)}_{du\nu e,1111}=C^{(9)}_{du\nu e,2111}=\frac{1}{\Lambda^2}$ or $C^{(9)}_{L\nu Qd,1111}=C^{(9)}_{L\nu Qd,1112}=\frac{1}{\Lambda^2}$ to induce $K^-\rightarrow\pi^+e^-e^-$. Similarly, we  set $ C^{(9)}_{H\nu e,12}$, $C^{(9)}_{Qu\nu L,1112}$,  $C^{(9)}_{du\nu e,1112}=C^{(9)}_{du\nu e,2112}$ and  $C^{(9)}_{L\nu Qd,2111}=C^{(9)}_{L\nu Qd,2112}$  to $\frac{1}{\Lambda^2}$ to induce   $K^-\rightarrow\pi^+\mu^-\mu^-$.

\begin{figure}[t]
		\includegraphics[scale=0.65]{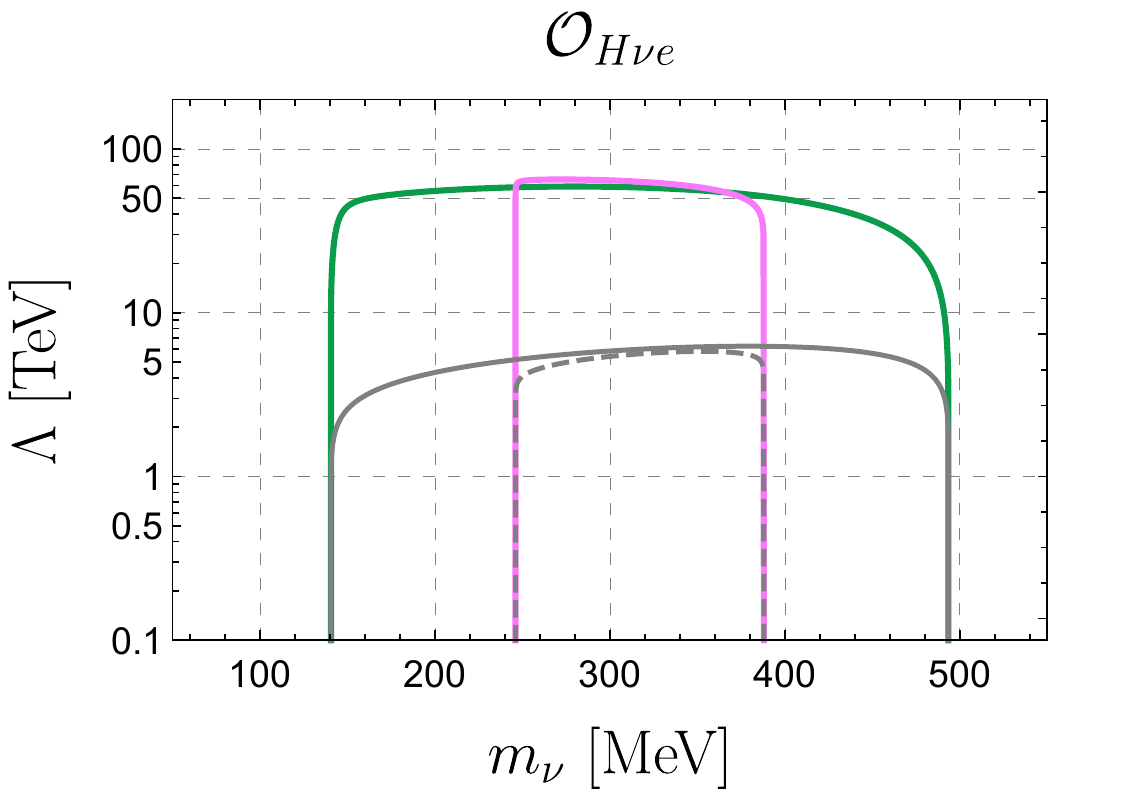}
		\includegraphics[scale=0.65]{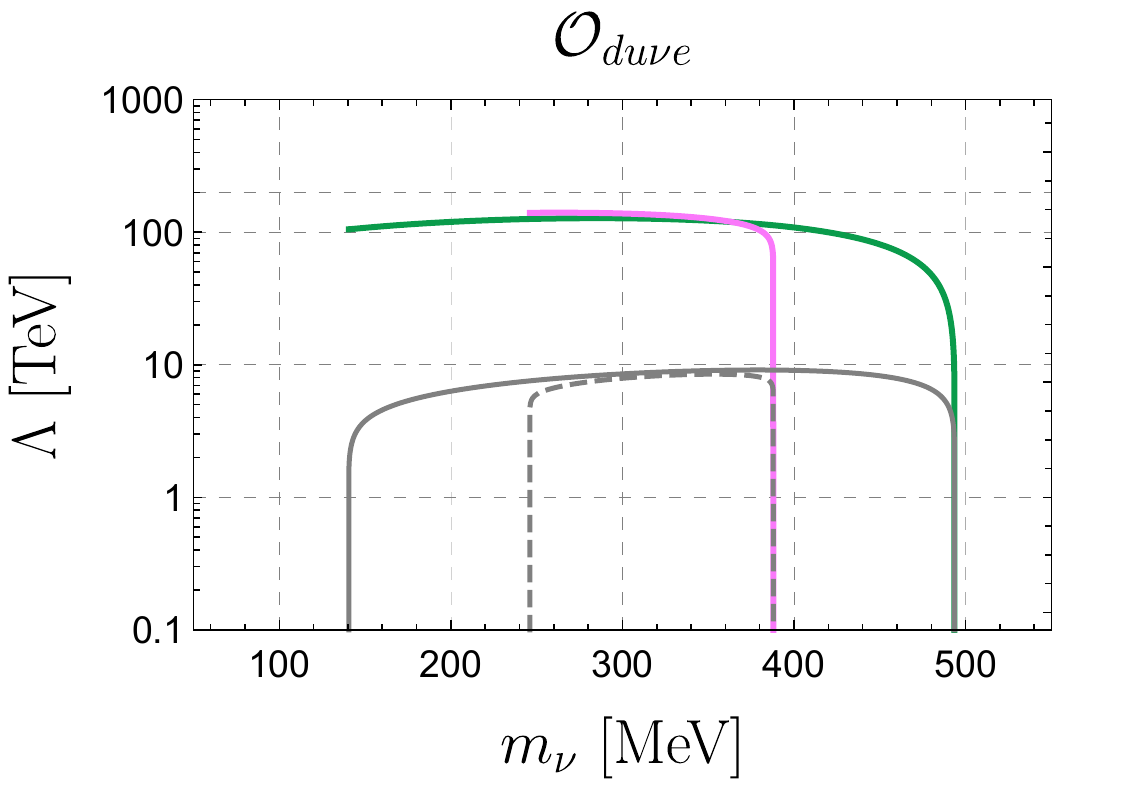}\\
		\includegraphics[scale=0.65]{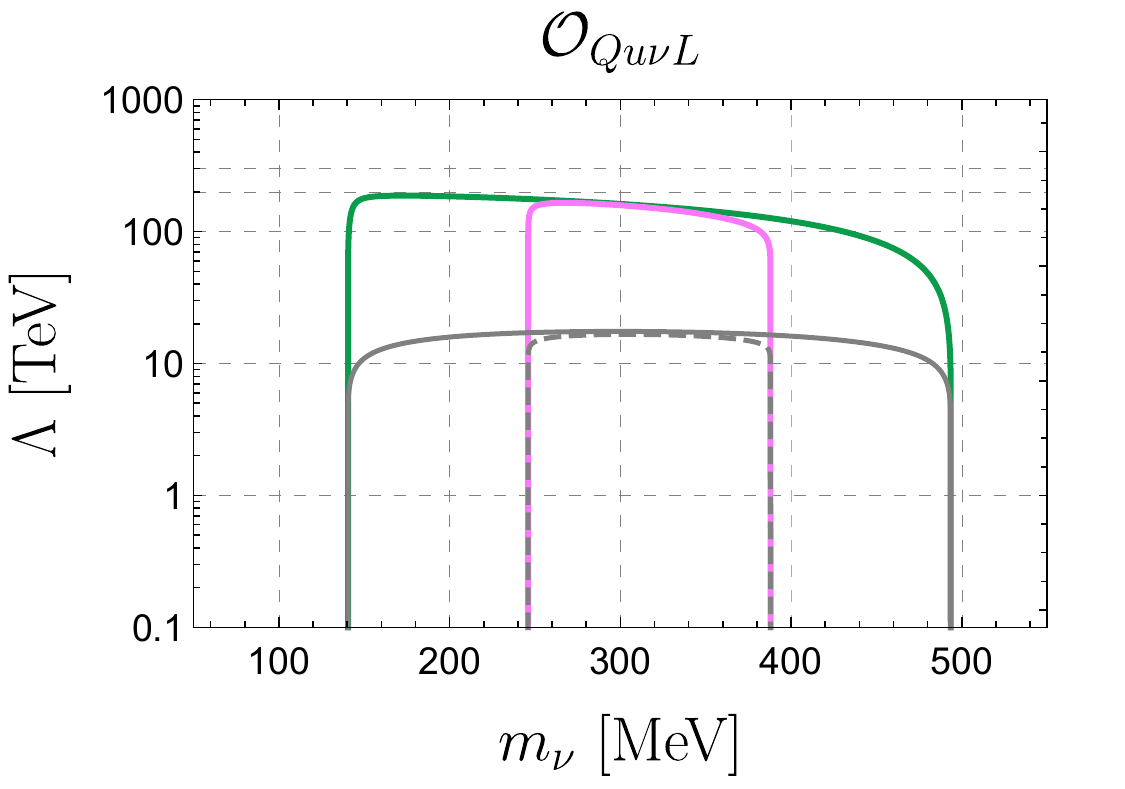}
		\includegraphics[scale=0.65]{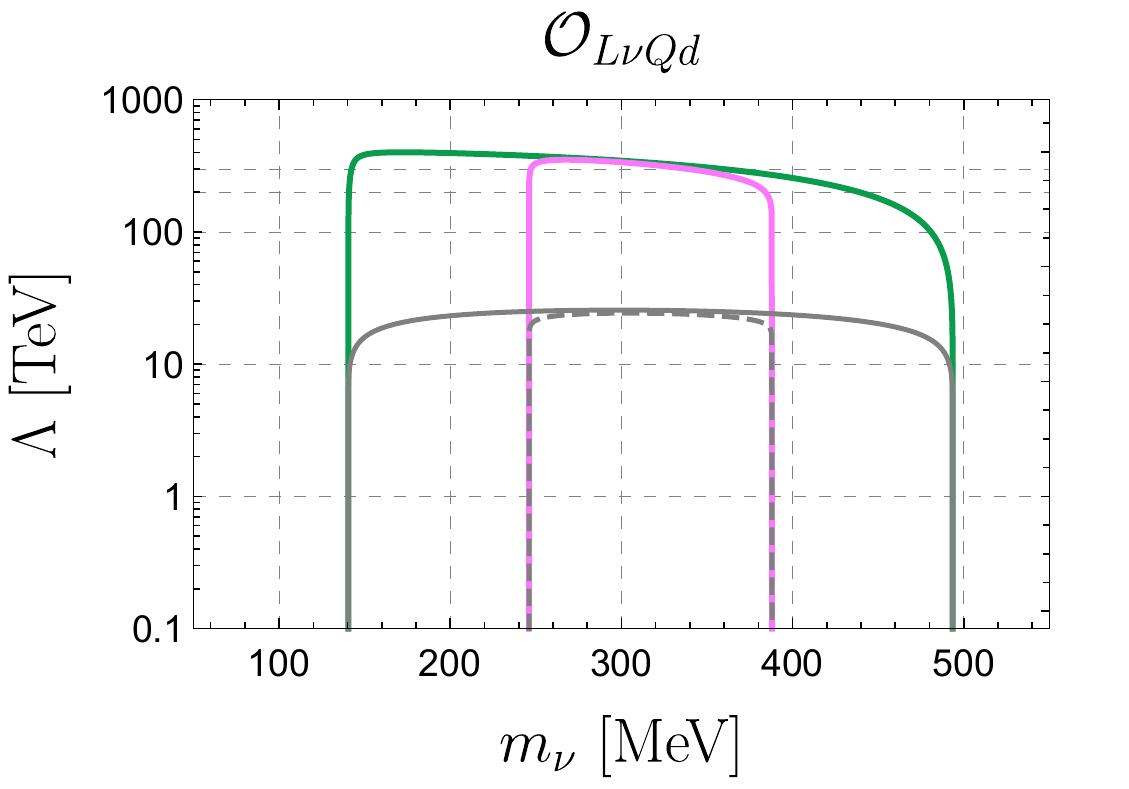}
		\caption{Same as Fig. \ref{fig:limitlq} but now we present limits on the BSM scale $\Lambda$ associated to various \textoverline{dim-6} operators.   }\label{fig:limitall}
\end{figure}
We show the limits on  $\Lambda$ for these four operators  in Fig. \ref{fig:limitall}. The two scalar-type operators give  stronger limits  than the two vector-type operators, because the decay rates from the latter are relatively suppressed by $m_\pi^2/B^2$. The contributions to the LNV kaon decay rate from $\mathcal O^{(9)}_{H\nu e}$ and  $\mathcal O^{(9)}_{Qu\nu L}$ are further suppressed by $|V_{us}|^2$. Note that near the threshold $m_\nu=m_l +m_\pi$ for the plot of $\mathcal O^{(9)}_{du\nu e}$ ,  the decay rates of the sterile neutrino and the  $|\vec{p}_2|$ in Eq. (\ref{eq:int}) approach to zero  at the same speed. Hence the two curves for  $\Lambda$ approach to some fixed values instead of  going down straightly around  $m_\nu=m_l +m_\pi$. While for the plots from other three operators, the decay rates of the sterile neutrino are not zero around $m_\nu=m_l +m_\pi$ as the sterile neutrino can still decay into light particles, e.g. $\pi^0+\nu_e$. We refer to Ref. \cite{Zhou:2021ylt,deVries:2020qns} for a more detailed discussion and calculation on the decay modes of the sterile neutrino for various \textoverline{dim-6} operators.
\subsection{Finite detector size effect}
In the resonance region where a sterile neutrino could be produced on-shell, we consider the intermediate neutrino as a real particle and it propagates for some distance before decaying into a pion and a charged lepton. In the case when sterile neutrinos decay outside the detector, we can not reconstruct the LNV process and thus get no valuable bound on the mixing angles or energy scales of the higher-dimensional operators. The observed branching ratio is suppressed due to the fact that some neutrinos decay outside the detector. Refs. \cite{Cvetic:2016fbv,Chun:2019nwi,Cvetic:2017vwl,Asaka:2016rwd,Zhang:2021wjj,Dib:2014iga} have discussed this effect. In this work we follow the method used in Ref. \cite{Chun:2019nwi} to make a rough estimate. We include the finite size detector effect by a probability factor $P_\nu$, which is the probability of $\nu_R$ to decay within the detector. We  write $P_\nu$ as 
\begin{equation}
P_\nu = 1- e^{-\frac{L_D}{L_\nu}}\,,
\end{equation}
where $L_D$ denotes the typical detector length and  $L_\nu=\frac{p_\nu}{m_\nu \Gamma_\nu}$ with $p_\nu$ the momentum of $\nu_R$. In the rest frame of $K^-$, the momentum of the sterile neutrino is given by
\begin{equation}
p^\star_\nu = \frac{\lambda^\frac{1}{2}(m_l\,,m_K\,,m_\nu)}{2m_K}\,.
\end{equation}
We can relate the energy $E_\nu$ of $\nu_R$ in the lab frame with those quantities in the rest frame of $K^-$ by the relation
\begin{equation}
E_\nu =E^\star_\nu (\gamma+ \frac{p^\star_\nu}{E^\star_\nu}\sqrt{\gamma^2-1}\cos \theta^\star_\nu)\,,
\end{equation} 
where $\gamma=\frac{E_K}{m_K}$ denoting the boost factor of $K^-$, $E^\star_\nu$ is the energy of $\nu_R$ in the rest frame of $K^-$ and $\theta^\star_\nu$ is the emission angle of $\nu_R$ relative to the velocity direction of $K^-$ in the rest frame of $K^-$. The energy of $\nu_R$ lies in the range $E^-_\nu<E_\nu<E^+_\nu$ with $E^\pm_\nu= \gamma E^\star_\nu \pm p^\star_\nu\sqrt{\gamma^2-1}$ and obeys a flat distribution.

We then can get the total number of LNV kaon decays inside the detector by integrating neutrino energy $E_\nu$
\begin{equation}
\begin{aligned}
N_\text{event} &= N_{K^-} \int_{E^-_\nu}^{E^+_\nu} dE_\nu \frac{\text{BR}(K^-\rightarrow l^- l^- \pi^+ )}{E^+_\nu-E^-_\nu} P_\nu\,,\\
&\approx N_{K^-} \int_{E^-_\nu}^{E^+_\nu} dE_\nu \frac{\text{BR}(K^-\rightarrow  l^- \nu_R )}{E^+_\nu-E^-_\nu}\frac{\Gamma(\nu_R\rightarrow l^-\pi^+)}{\Gamma_\nu} P_\nu\,,
\end{aligned}
\end{equation}
where $N_{K^-}$ is the number of $K^-$.  In NA62 experiment, 400 GeV protons collide the target and produce a large number of $K^+$ mesons, which carry a momentum of 75 GeV. Assuming three years of running, the expected number of $K^+$ decays in the fiducial volume is $N_{K^+}=1.35\times 10^{13}$. 
Following Refs. \cite{Feldman:1997qc,Chun:2019nwi}, we assume zero background events and $L_D\approx 65$ m. 

By requiring the signal events to be $N_\text{event}=3.09$ we  get bound on the mixing angle or new physics energy scale $\Lambda$ as a function of sterile neutrino mass $m_\nu$. We show our results in Figs. \ref{fig:limitSM}-\ref{fig:limitall}, where we use the gray lines to denote the sensitivity for $K^-\rightarrow\pi^+ e^- e^-$ and gray dashed lines for $K^-\rightarrow\pi^+ \mu^- \mu^-$. Through the gray (dashed) lines, we find the limits on $|U^2_{l4}|$ are of the order  $10^{-6}$, and the limits on $\Lambda$  are 5 - 30 TeV.\footnote{In principle one should also consider  the  decay length of $K^+$ and  the geometry of  NA62 experiment.  We leave this careful analysis in the future work.}

In the minimal scenario the gray (dashed) lines we get in Fig. \ref{fig:limitSM} are above the  region excluded by big bang nucleosynthesis (BBN) \cite{Sabti:2020yrt,Boyarsky:2020dzc}. However they are slightly weaker than the limits from \textbf{ATLAS} experiment \cite{deVries:2020qns}, and weaker than the constraints from $K^+\rightarrow l^+\nu_R$ \cite{NA62:2020mcv,NA62:2021bji} by two to three order of magnitude. This makes sense as we  require additionally $\nu_R$ to decay within the detector.

The constraints on the higher-dimensional operators mentioned in previous subsection have been probed a lot in Refs. \cite{Li:2020lba,Li:2020wxi,Alcaide:2019pnf,Biekotter:2020tbd,Mandal:2020htr,COHERENT:2017ipa,Han:2020pff,Bischer:2019ttk} via elastic coherent neutrino-nucleon scattering, missing transverse energy searches, lepton flavor universality, CKM unitarity, meson and tau decays, etc. For the sterile neutrino with mass $m_\pi < m_\nu <m_K$, the bounds they got are from 1 TeV to 10 TeV, which are  weaker than those from LNV kaon decays. Ref. \cite{deVries:2020qns} investigated higher-dimensional operators via displaced vertices search for the sterile neutrino at the LHC and \textbf{SHiP}, which could  reach a limit 20 - 30 TeV.  Neutrinoless double-beta decay \cite{Dekens:2020ttz} gives a stronger limit roughly 50 TeV.   Despite the different flavor configurations and the narrow parameter space of the sterile neutrino mass, our results    are close and complementary to their results.

\section{Conclusions}\label{sec:conclusion}
 In this work, we  used a systematic framework to study  the effect of light sterile neutrinos with mass smaller than the electroweak scale, $m_\nu<v$, on  the lepton-number-violating decays $K^-\rightarrow \pi^+ l^- l^-$.  
 The sterile neutrinos $N$ are gauge singlets under the SM gauge group and are allowed to interact with the SM fields through the Yukawa interaction and gauge-invariant higher-dimensional interactions up to \textoverline{dim-7}. After integrating out heavy SM particles, we  match the $\nu$SMEFT operators onto $SU(3)\times U(1)_{em}$-invariant operators. If the sterile neutrino mass is above $\Lambda_\chi$, we also integrate it out and get dim-9 operators, which induce the short distance contributions to LNV kaon decays.   For the sterile neutrino with a mass $m_\nu <\Lambda_\chi$, we get LNV operators of  dim-6 and dim-7 in Eqs. (\ref{eq:lowenergy6_l2}) and  (\ref{eq:lowenergy7}), and LNC operators in Eqs. (\ref{eq:lowenergy6_l0}) and (\ref{eq:lowenergy7b}). 
  By using  chiral perturbation  theory, we connect dim-6, -7 and -9 operators   at the quark level to mesonic physics and  give the expressions for the decay amplitude, which includes the   long-distance (from dim-6 and -7 operators) or short-distance contributions  depending  on whether the sterile neutrino mass is above or below $\Lambda_\chi$.
  
  We find the presence of a sterile neutrino and the non-standard interactions have a dramatic impact on the LNV kaon decay phenomenology. In the case of a sterile neutrino with mass $\Lambda_\chi<m_\nu<v$,   the  new physics scale  probed by the dim-9 operators is improved by one order compared  to the case without a sterile neutrino \cite{Liao:2019gex}. Nevertheless, the BSM physics scale that can be probed is of order $\mathcal{O}$(100) GeV which is still too low to make the $\nu$SMEFT approaches valid. The limits are also much weaker than the corresponding $0\nu\beta\beta$ decay limits for similar operators with different generation indices because of the relatively small data samples of kaon experiments. However,  very stringent bounds on the EFT operators emerge if the sterile neutrino mass lies in the resonance region $(m_\pi+m_l,m_K-m_l)$. The resulting lepton-number-violating decay rate are highly enhanced and we obtain strong limits on the neutrino mixing angles $|U^2_{e4}|$ and $|U^2_{\mu4}|$ at the level of $10^{-9}$-$10^{-10}$, close to seesaw predictions, and on the BSM scales $\Lambda$, up to $\mathcal O$(300) TeV, for various higher-dimensional operators. After considering the finite detector size effect, we find the limits on mixing angle $|U^2_{l4}|$ become at the level of  $10^{-6}$ and  the BSM scales $\Lambda$ are weakened to  $\mathcal O$(30) TeV. These limits obtained this way are very strong though only in a narrow window of sterile neutrino masses with 150 MeV$<m_\nu<$490 MeV.  The framework we developed in this work can be readily extended  to probe BSM physics in other types of  decays, e.g. the  LNV decays of charm and bottom mesons, or $\tau$ leptons.

\medskip
\section*{Acknowledgements}
I thank  Jordy de Vries for valuable discussions and suggestions  on this work and for his  precious  help and feedback in writing. I also thank Wouter Dekens for his nice comments and Xiaodong Ma for useful discussions. 

\appendix
\section{Matching at the EW scale}\label{app:matching}
We give the explicit matching conditions for all the operators in Eq.~\eqref{DL2lag}. For the mass terms we have 
\bea\label{Massmatch}
M_L &=& -v^2 C^{(5)}-\frac{v^4}{2} C_{LH}\,,\nn\\
M_R &=& \bar M_R + v^2 \bar M_R^{(5)}-\frac{v^4}{2} C_{\nu H}^{(7)}\,,\nn\\
M_D &=&\frac{v}{\sqrt{2}} \left[Y_\nu -\frac{v^2}{2}C_{L\nu H}^{(6)}\right]\,.
\eea 
The matching conditions for dim-6 operators involving active neutrinos $\nu_L$ are 
\bea\label{match6LNC}
c_{\rm VL}^{(6)} &=& -2\mathbb{1}V+2v^2\left[C_{LQ\,3}^{(6)}-C_{HL\,3}^{(6)}-C_{HQ\,3}^{(6)}\, \mathbb{1}\right]V-\frac{4\sqrt{2}v}{g }M_e \left(C^{(6)}_{eW}\right)^\dagger V\nn\\
&&-\frac{4\sqrt{2}v}{g} C^{(6)}_{\nu W} M_D^\dagger V+4v^2 \left(C_{LHW}^{(7)}\right)^\dagger M_L V\,,\nn\\
c_{\rm VR}^{(6)} &=&-v^2C_{Hud}^{(6)}\, \mathbb{1},\nn\\
c_{\rm SR}^{(6)} &=& v^2 \left(C_{LedQ}^{(6)}\right)^\dagger\,,\nn\\
c_{\rm SL}^{(6)} &=& v^2\left(C_{LeQu\,1}^{(6)}\right)^\dagger V\,,\nn\\
c_{\rm T}^{(6)} &=& v^2\left(C_{LeQu\,3}^{(6)}\right)^\dagger V\,,\nn\\
\frac{1}{v^{3}}\, C^{(6)}_{\textrm{VL},ij} & = & - \frac{i}{\sqrt{2}}    C_{LHDe,ji}^{(7)\,*}V + 4   \frac{m_e}{v}  C_{LHW,ji}^{(7)\,*}V -\frac{4\sqrt{2}}{gv^2}\left(M_L C_{eW}^{(6)}\right)^*_{ji}V +\frac{8}{gv}\left(M_D C_{\nu e W}^{(7)\,*}\right)_{ji}V
\,, \nonumber \\ 
\frac{1}{v^{3}}\,C^{(6)}_{\textrm{VR},ij} &=& \frac{1}{\sqrt{2}}   C_{Leu\bar dH,ji}^{(7)\,*} \,,  \nonumber \\
\frac{1}{v^{3}}\,C^{(6)}_{\textrm{SR},ij} & =& \frac{1}{2\sqrt{2}} \left( C^{(7)}_{LL Q \bar d H\,2,ij} - C^{(7)}_{LL Q \bar d H\,2,ji}+  C^{(7)}_{LL Q \bar{d} H\,1 ,ij} \right)^* \nn\\
&&+ \frac{V_{ud}}{2}\frac{m_d}{v} \left( C^{(7)}_{LHD\,1,ij}- C^{(7)}_{LHD\,1,ji}- C^{(7)}_{LHD\,2,ji} \right)^*
-\frac{i}{2}\frac{m_u}{v}\left( C^{(7)}_{LL \bar d u D\,1,ij}- C^{(7)}_{LL \bar d u D\,1,ji}\right)^*\,, \nonumber \\  
\frac{1}{v^{3}}\,C^{(6)}_{\textrm{SL},ij}  &=&  
\frac{1}{\sqrt{2}}   C_{LL \bar Q u H,ij}^{(7)\,*}V+\frac{1}{2v} \left[\left(C_{QL\nu uD}^{(7)}\right)^\dagger M_D^T\right]_{ij}V \nn\\
&&- \frac{V_{ud}}{2}\frac{m_u}{v} \left( C^{(7)}_{LHD\,1,ij}- C^{(7)}_{LHD\,1,ji}-    C^{(7)}_{LHD\,2,ji} \, \right)^*V
+\frac{i}{2}\frac{m_d}{v}\left(  C^{(7)}_{LL \bar d u D\,1,ij}- C^{(7)}_{LL \bar d u D\,1,ji}\right)^*V\,,  \nonumber \\
\frac{1}{v^{3}}\,C^{(6)}_{\textrm{T},ij } &= &  \frac{1}{8 \sqrt{2}} \left(C^{(7)}_{LL Q \bar d H\,2,ij}+ C^{(7)}_{LL Q \bar d H\,2,ji}+  C^{(7)}_{LL Q \bar d H\,1,ij}\right)^* \,,
\eea
where  the indices $ij$ denote the generation of leptons and the indices  of quarks are implied.
While for the dim-6 operators with sterile neutrinos $\nu_R$, we have
\bea\label{match6LNCsterile}
\bar c_{\rm VL}^{(6)} &=& \left[-v^2C_{H\nu e}^{(6)}+\frac{8v^2}{g} M_R^\dagger C_{\nu eW}^{(7)} - \frac{4\sqrt{2} v}{g}   \left(C_{\nu W}^{(6)}\right)^\dagger M_e-\frac{4\sqrt{2}v}{g}M_D^\dagger C_{eW}^{(6)} \right]^\dagger V\,,\nn\\
\bar c_{\rm VR}^{(6)} &=& v^2\left(C_{du\nu e}^{(6)}\right)^\dagger\,,\nn\\
\bar c_{\rm SR}^{(6)}&=& -v^2C_{L\nu Qd}^{(6)}+\frac{v^2}{2} C_{LdQ\nu }^{(6)}\,,\nn\\
\bar c_{\rm SL}^{(6)}&=& v^2\left(C_{Qu\nu L}^{(6)}\right)^\dagger V+\frac{v^2}{2} \left(C_{QL\nu uD}^{(7)}\right)^\dagger M_R V\,,\nn\\
\bar c_{\rm T}^{(6)} &=& \frac{v^2}{8} C_{LdQ\nu }^{(6)}\,,\nn\\
\bar C_{\rm VL}^{(6)} &=&-\frac{4\sqrt{2} v}{g}C_{\nu W}^{(6)} M_R^\dagger V+\frac{v^3}{\sqrt{2}} C_{\nu L1}^{(7)} V+ 8 \frac{v^2}{g} \left(C_{\nu eW}^{(7)}\right)^\dagger M_e V\nn\\
&&+\left(\frac{v}{\sqrt{2}}\right)^3 \left(C_{Q\nu QLH2}^{(7)}\right)^\dagger V+4v^2 \left(C_{LHW}^{(7)}\right)^\dagger M_D^* V\,,\nn\\
\bar C_{\rm VR}^{(6)} &=& -\frac{v^2}{2} m_d \left(C_{QL\nu uD}^{(7)}\right)^\dagger+\left(\frac{v}{\sqrt{2}}\right)^3 \left(C_{dL\nu uH}^{(7)}\right)^\dagger\,,\nn\\
\bar C_{\rm SR}^{(6)} &= & \left[\frac{v^3}{\sqrt{2}} C_{dQ\nu eH}^{(7)} +\frac{v^2}{2} M_e^\dagger C_{d\nu QLD}^{(7)} -\frac{v^2}{2} m_d C_{\nu eD}^{(7)}-\frac{v^2}{2} m_u C_{du\nu eD}^{(7)}\right]^\dagger\,,\nn\\
\bar C_{\rm SL}^{(6)} &=& \left[\frac{v^3}{\sqrt{2}} C_{Qu\nu eH}^{(7)}-\left(\frac{v}{\sqrt{2}}\right)^3 C_{Qe\nu uH}^{(7)} + \frac{v^2}{2} m_d C_{du\nu eD}^{(7)}+\frac{v^2}{2}  C_{QL\nu uD}^{(7)}M_e +\frac{v^2}{2} m_u C_{\nu eD}^{(7)}\right]^\dagger V\,,\nn\\
\bar C_{\rm T}^{(6)} &=& -\frac{v^3}{8\sqrt{2}} \left(C_{Qe\nu uH}^{(7)}\right)^\dagger V + \frac{v^2}{8} M_e^\dagger \left(C_{QL\nu uD}^{(7)}\right)^\dagger V\,.
\eea
The matching conditions of dim-7 operators can be obtained from \cite{Dekens:2020ttz} and we ignore them here as they are not important in this work. The matching conditions for the dim-9 operators can be taken from Ref.~\cite{Cirigliano:2017djv,Liao:2019gex}
\bea\label{match9}
\frac{1}{v^{3}}\,C^{(9)}_{1L} &=& - 4 V_{ud} V_{us} \left(   C^{(7)}_{LHD\, 1}+ 4 \mathcal C_{LHW}\right)^*\,, \nn\\
\frac{1}{v^{3}}\,C^{(9)}_{5L} &=&  4 i V_{ud} \,    C^{(7)*}_{LL \bar d u D \, 1,us}\,,\nn\\
\frac{1}{v^{3}}\,C^{(9)\prime}_{5L} &=&  4 i V_{us} \,    C^{(7)*}_{LL \bar d u D \, 1,ud}\,.
\eea
\section{Additional contributions to the dim-9 operators}\label{app:matchd9}
In general, four-quark two-lepton operators with an additional derivative  are also induced when integrating out a heavy neutrino. When we  match them onto the Chiral Perturbation Theory,  a lot of new LECs arise. In table \ref{tab:matchingd9} we  give the matching conditions only for  interactions, which via the equations of motions can be written as $m_q \times \mathcal O^{(9)}$ or $m_l \times \mathcal O^{(9)}$ with $m_q$ being the light quark mass and $m_l$ the  mass of charged lepton.  The remaining  terms  contain a derivative and are of dim-10, which result in unknown LECs when matched onto Chiral Perturbation Theory. Thus we neglect them here. 
 To make the expressions in a compact form, we have removed an overall factor $\frac{1}{v^4}\frac{1}{m^2_\nu}$ and the dim-6 WCs,  and we also use  $E_\mu =\bar{e}\gamma_\mu\gamma_5 C \bar{e}^T$ and $E_{L, R}=\bar{e}_{L, R}C\bar{e}^T_{L, R}$.

{\renewcommand{\arraystretch}{1.3}\begin{table}[t!]\small
		\center
		\scalebox{0.87}{\begin{tabular}{|c|c|c|c|c|c|}
			\hline
			& $C_{\rm VLR,ud}^{(6)}$  & $C_{\rm VRR,ud}^{(6)}$  & $C_{\rm SRR,ud}^{(6)}$  & $C_{\rm SLR,ud}^{(6)}$  &$C_{\rm TRR,ud}^{(6)}$ 
\\ \hline
			$C_{\rm VLL,us}^{(6)}$ &$\frac{1}{2}m_d \mathcal O^\mu_{6, \rm usud}E_\mu$ & $\frac{1}{2}m_d \mathcal O^\mu_{8, \rm usud}E_\mu$& $m_u \mathcal O^\prime_4E_L$&$m_u \mathcal O_2E_L$& $8m_u\mathcal  O^\prime_5 E_L$
\\
			& $-\frac{1}{2}m_u \mathcal O^\mu_{8, \rm usud} E_\mu$&$-\frac{1}{2}m_u \mathcal O^\mu_{6, \rm usud} E_\mu$&$-m_s \mathcal O^\prime_2E_L$& $-m_s \mathcal O_4E_L$&  $-4m_d\mathcal  O_1 E_L$
			
\\
		&	+$m_l \mathcal O_1 E_L$&$-2m_l \mathcal O^\prime_5 E_L$&$-\frac{1}{2}m_l \mathcal O^\mu_{6,\rm usud}E_\mu$&$-\frac{1}{2}m_l \mathcal O^\mu_{8,\rm usud}E_\mu$&
			
\\ \hline
$C_{\rm VRL,us}^{(6)}$  &$\frac{1}{2}m_d \mathcal O^{\mu\prime}_{8, \rm usud}E_\mu$ & $\frac{1}{2}m_d \mathcal O^{\mu\prime}_{6, \rm usud}E_\mu$& $m_u \mathcal O^\prime_2E_L$&$m_u \mathcal O_4E_L$& $8m_d\mathcal  O_5 E_L$
			\\
			& $-\frac{1}{2}m_u \mathcal O^{\mu\prime}_{6, \rm usud} E_\mu$&$-\frac{1}{2}m_u \mathcal O^{\mu\prime}_{8, \rm usud} E_\mu$&$-m_s \mathcal O^\prime_4E_L$& $-m_s \mathcal O_2E_L$&  $-4m_u\mathcal  O^\prime_1 E_L$
			
			\\
			&	$-2m_l \mathcal O_5 E_L$&$+m_l \mathcal O^\prime_1 E_L$&$-\frac{1}{2}m_l \mathcal O^{\mu\prime}_{8,\rm usud}E_\mu$&$-\frac{1}{2}m_l \mathcal O^{\mu\prime}_{6,\rm usud}E_\mu$&

\\ \hline
$C_{\rm SRL,us}^{(6)}$ & $-m_d \mathcal O^\prime_2 E_R$  & $-m_d \mathcal O_4 E_R$   & $m_l \mathcal O^\prime_2 E_R$  & $m_l \mathcal O_4 E_R$  & $-2 m_d \mathcal O^\mu_{6, \rm udus}E_\mu$
		
\\
    & $+m_u\mathcal O_4 E_R$ &$+m_u\mathcal O^\prime_2 E_R$   &  &  &		 $-2 m_u \mathcal O^{\mu\prime}_{8, \rm udus}E_\mu$
\\    
   &  $\frac{1}{2}m_l\mathcal O^\mu_{6, \rm udus} E_\mu$& $\frac{1}{2}m_l\mathcal O^{\mu\prime}_{8, \rm udus} E_\mu$ &  &  &

\\ \hline
			$C_{\rm SLL,us}^{(6)}$ &  $-m_d \mathcal O^\prime_4 E_R$  & $-m_d \mathcal O_2 E_R$   & $m_l \mathcal O^\prime_4 E_R$  & $m_l \mathcal O_2 E_R$  & $-2 m_d \mathcal O^\mu_{8, \rm udus}E_\mu$
			
			\\
			& $+m_u\mathcal O_2 E_R$ &$+m_u\mathcal O^\prime_4 E_R$   &  &  &		 $-2 m_u \mathcal O^{\mu\prime}_{6, \rm udus}E_\mu$
			\\    
			&  $\frac{1}{2}m_l\mathcal O^\mu_{8, \rm udus} E_\mu$& $\frac{1}{2}m_l\mathcal O^{\mu\prime}_{6, \rm udus} E_\mu$ &  &  & 
			
		\\ \hline
			$C_{\rm TLL,us}^{(6)}$ & $-4 m_u \mathcal O_1 E_R$&$-4 m_s \mathcal O^\prime_1 E_R$&$2m_s \mathcal O^\prime_{8, \rm usud} E_\mu$&$2m_u \mathcal O_{8, \rm usud} E_\mu$& $-8 m_s(2 \mathcal O^\prime_{9, \rm udus}+\mathcal O^\prime_{8, \rm usud})E_\mu$
			\\
			 & $+8 m_s \mathcal O_5 E_R$& $+8 m_u \mathcal O^\prime_5 E_R$&  $+2m_u \mathcal O_{6, \rm usud} E_\mu$ &$+2m_s \mathcal O^\prime_{6, \rm usud} E_\mu$& $+8 m_u(2 \mathcal O_{7, \rm usud}+\mathcal O_{6, \rm usud})E_\mu$
		\\ \hline
		\end{tabular}}
		\caption{The dim-9 interactions induced by integrating a heavy neutrino between two dim-6 operators involving neutrinos of  different chiralities (the coefficients have been divided by $\frac{1}{v^4}\frac{1}{m^2_\nu}$ and the corresponding two dim-6 WCs).  }  \label{tab:matchingd9}
\end{table}}

In principle dim-9 interactions are also induced by  terms, $e.g.$  $C_{\rm VLR,us}^{(6)}\times C_{\rm VLL,ud}^{(6)}$. We can easily get the dim-9 interactions induced by   $C_{\rm VLR,us}^{(6)}\times C_{\rm VLL,ud}^{(6)}$ via a replacement $d\leftrightarrow s$ on the dim-9 interactions induced by  $C_{\rm VLL,us}^{(6)}\times C_{\rm VLR,ud}^{(6)}\,$.

\section{Sterile neutrino decay processes}\label{app:decay}
In this section we discuss  possible decay modes of the sterile neutrino with a mass $m_\nu$ in the resonance region $(m_\pi+m_l, m_K-m_l)$ and give their expressions in  analytical forms. 
\subsection{Decay modes in the minimal scenario}
In the minimal scenario we find the decay rate for  $\nu_R\rightarrow l^\mp \pi^\pm$ is , 
  \begin{equation}
  \begin{aligned}
 \Gamma (\nu_R\rightarrow l^\mp \pi^\pm) &= 2\times \frac{\sqrt{\lambda(m_\nu,m_l, m_\pi)}}{8 \pi m^3_\nu}G^2_F F^2_0 |V_{ud}|^2 |U_{l4}|^2  \\
  &\times ((m_\nu^2-m_l^2)^2-m^2_\pi(m^2_l+m^2_\nu))\theta(m_\nu-m_l-m_\pi)\,,
  \end{aligned}
  \end{equation}
where $l=e, \mu$ and we add a 2 to account for the Majorana nature. The decay rate of $\nu_R \rightarrow \nu_l \pi^0$ is given by \cite{deVries:2020qns}
\begin{equation}
 \Gamma (\nu_R\rightarrow\nu_l \pi^0)= 2\times \frac{G^2_F F^2_0m^3_\nu |U_{l4}|^2}{16\pi}(1-\frac{m^2_{\pi^0}}{m^2_\nu})^2 \theta(m_\nu-m_{\pi^0})\,.
\end{equation}
The sterile neutrino can also decay into three light active neutrinos and the decay rates are \cite{Bondarenko:2018ptm}
\begin{equation}
\Gamma (\nu_R \rightarrow\nu_\alpha \nu_\beta\bar{\nu}_\beta) =2\times (1+\delta_{\alpha \beta}) \frac{G_F^2 m^5_\nu |U_{\alpha 4}|^2}{768\pi^3}\,,
\end{equation}
where $\alpha=e, \mu$ and $\beta=e, \mu, \tau$ are the flavor indices of the active neutrinos. The three-body decay rates for sterile neutrino into two charged leptons and   one active neutrino can not be written analytically. Thus, we use the method of Ref. \cite{deVries:2020qns} and use \texttt{FeynCalc}~\cite{Shtabovenko:2020gxv,Shtabovenko:2016sxi,Mertig:1990an} to do the phase space integrals numerically.

\subsection{Decay modes in the leptoquark scenario}
Without considering the interactions in the minimal scenario, there are only two types of decay modes for the sterile neutrino in the leptoquark scenario. The decay rate for $\nu_R\rightarrow l^\mp\pi^\pm$ is 
\begin{equation}
\Gamma (\nu_R\rightarrow l^\mp\pi^\pm) = 2\times \frac{\sqrt{\lambda(m_\nu,m_l,m_{\pi})}}{32\pi m_\nu^3}(\frac{v^2}{2 m^2_{\rm LQ}})^2 G^2_F F^2_0 B^2 (m^2_\nu-m^2_\pi+m^2_l)\theta(m_\nu-m_l-m_\pi)\,,
\end{equation}
and for  $\nu_R\rightarrow \nu_l \pi^0$ we find

\begin{equation}
\Gamma (\nu_R\rightarrow \nu_l \pi^0) = 2\times \frac{1}{64\pi m^3_\nu}(\frac{v^2}{2 m^2_{\rm LQ}})^2 G^2_F F^2_0 B^2 (m^2_\nu-m^2_{\pi^0})^2 \theta(m_\nu-m_{\pi^0})\,.
\end{equation}

\bibliographystyle{utphysmod}
\bibliography{bibliography}

\end{document}